\documentclass[journal]{IEEEtai}
\usepackage[dvipsnames]{xcolor}
\usepackage{amsmath,amsfonts}
\usepackage{algorithmic}
\usepackage{algorithm}
\usepackage{array}
\usepackage{amssymb}
\usepackage{textcomp}
\usepackage{stfloats}
\usepackage{url}
\usepackage{verbatim}
\usepackage{graphicx}
\usepackage{cite, hyperref}
\usepackage{comment}
\usepackage{caption}
\usepackage{subcaption}
\usepackage{multirow}
\usepackage{xcolor}
\usepackage{soul}
\usepackage[]{mdframed}
\usepackage[normalem]{ulem}
\usepackage{float}
\usepackage{times,epsfig}
\usepackage{amsthm}

\begin{document}

\title{Introducing Chunk as a Service: A New Paradigm for Cost-Efficient RAG-Based LLM Systems}
\title{Chunk as a Service: A New Paradigm for Cost-Efficient RAG-Based LLM Systems with Budget-Constrained Variants}
\title{Online Limited-Budget RAG-Based LLM: A Chunk as a Service System Optimizes Performance at Lower Costs}
\title{Online Limited-Budget RAG-Based LLM System: Optimizing Performance at Lower Costs}
\title{Budget-Constrained Retrieval-Augmented Generation: The Chunk-as-a-Service Model}
\title{Budget-Constrained Online RAG-Based LLM:\\ A Chunk-as-a-Service Approach}
\title{Budget-Constrained Online RAG-Based LLM:\\ The Chunk-as-a-Service Model}
\title{Budget-Constrained Online Retrieval-Augmented Generation: The Chunk-as-a-Service Model}

 \author{Shawqi Al-Maliki, Ammar Gharaibeh, Mohamed Rahouti, \IEEEmembership{Member, IEEE}, Mohammad Ruhul Amin, Mohamed Abdallah, \IEEEmembership{Senior Member, IEEE}, Junaid Qadir, \IEEEmembership{Senior Member, IEEE}, Ala Al-Fuqaha$^*$, \IEEEmembership{Senior Member, IEEE}
    
    \thanks{S. Al-Maliki, M. Abdallah., and A. Al-Fuqaha are with the Information and Computing Technology (ICT) Division, College of Science and Engineering, Hamad Bin Khalifa University, Doha 34110, Qatar (email: \{shalmaliki, moabdallah, and aalfuqaha\}@hbku.edu.qa)}
    \thanks{A. Gharaibeh with School of Electrical Engineering and Information Technology, German Jordanian University, Amman, Jordan (email: ammar.gharaibeh@gju.edu.jo)}
    \thanks{M. Rahouti and M. Ruhul are with Department of Computer and Information Sciences, Fordham University, Bronx, NY, USA (email: mrahouti@fordham.edu and mamin17@fordham.edu)}
    \thanks{J. Qadir is with the Department of Computer Science and Engineering, College of Engineering, Qatar University, Doha, Qatar (email: jqadir@qu.edu.qa)}
    \thanks{Corresponding author: Ala Al-Fuqaha (Email: aalfuqaha@hbku.edu.qa)}}
    

\maketitle

\begingroup
\renewcommand\thefootnote{}
\footnotetext{© 2026 IEEE.  Personal use of this material is permitted.  Permission from IEEE must be obtained for all other uses, in any current or future media, including reprinting/republishing this material for advertising or promotional purposes, creating new collective works, for resale or redistribution to servers or lists, or reuse of any copyrighted component of this work in other works.}
\endgroup
    
\begin{abstract}
Large Language Models (LLMs) have revolutionized the field of natural language processing. However, they exhibit some limitations, including a lack of reliability and transparency: they may hallucinate and fail to provide sources that support the generated output. Retrieval-Augmented Generation (RAG) was introduced to address such limitations in LLMs. 
One popular implementation, RAG-as-a-Service (RaaS), has shortcomings that hinder its adoption and accessibility.
For instance, RaaS pricing is based on the number of submitted prompts, without considering whether the prompts are enriched by relevant chunks, i.e., text segments retrieved from a vector database, or the quality of the utilized chunks (i.e., their degree of relevance). This results in an opaque and less cost-effective payment model.
We propose Chunk-as-a-Service (CaaS) as a transparent and cost-effective alternative. CaaS includes two variants: Open-Budget CaaS (OB-CaaS) and Limited-Budget CaaS (LB-CaaS), which is enabled by our ``Utility-Cost Online Selection Algorithm (UCOSA)''. UCOSA further extends the cost-effectiveness and the accessibility of the OB-CaaS variant by enriching, in an online manner, a subset of the submitted prompts based on budget constraints and utility-cost tradeoff.
Our experiments demonstrate the efficacy of the proposed UCOSA compared to both offline and relevance-greedy selection baselines.
In terms of the performance metric---the number of enriched prompts ($NEP$) multiplied by the Average Relevance ($AR$)---UCOSA outperforms random selection by approximately 52\% and achieves around 75\% of the performance of offline selection methods. 
Additionally, in terms of budget utilization, LB-CaaS and OB-CaaS achieve higher performance-to-budget ratios of 140\% and 86\%, respectively, compared to RaaS, indicating their superior efficiency.
\end{abstract}

\begin{IEEEImpStatement}
RAG systems help mitigate hallucinations in LLMs, and RAG-as-a-Service (RaaS) facilitates their broader utilization. However, the RaaS model lacks transparency in usage pricing, is less cost-effective, and is inaccessible to users with limited budgets. These shortcomings have clear negative social and economic impacts on potential users. Our proposed Chunk-as-a-Service (CaaS) model is a transparent and cost-effective variant of RaaS. It calculates usage pricing based on the actually utilized chunks during prompt enrichment, adhering to transparency as a key social principle. CaaS also introduces a new publishing paradigm aimed at creating competitive content and fair profits for content creators, CaaS providers, and end users. Furthermore, CaaS offers a Limited-Budget (LB-CaaS) variant, enabling users with limited budgets to access RAG, effectively addressing the economic barrier inherent in the traditional RaaS model.
\end{IEEEImpStatement}


\section{Introduction}
\label{sec:Introduction}

\IEEEPARstart{L}arge Language Models (LLMs) \cite{brown2020language, achiam2023gpt, reid2024gemini,chowdhery2023palm} have revolutionized Natural Language Processing (NLP) by enabling human-like text generation and comprehension \cite{yang2024harnessing}. These models have advanced rapidly, showcasing significant improvements in understanding and producing natural language. However, despite their impressive capabilities, LLMs often generate inaccurate outputs, or `hallucinations' \cite{ji2023survey, huang2023survey, zhang2024how}.
This problem becomes particularly evident when LLMs respond to prompts referencing information not present in their training data. Lacking the necessary context, LLMs may refuse to answer, produce incorrect responses, or hallucinate, resulting in erroneous or misleading outputs.

Retrieval-Augmented Generation (RAG) systems \cite{guu2020retrieval, lewis2020retrieval} address the limitations of LLMs by utilizing external, non-parameterized indexed data sources, such as knowledge bases. RAG systems enable seamless access to up-to-date and reliable information, significantly improving the efficiency and accuracy of the generated output. By integrating real-time retrieval of relevant data, RAG systems effectively mitigate the issue of LLM hallucinations, ensuring that responses are both current and contextually accurate. This approach enhances LLMs' overall reliability and usefulness, making them more robust and dependable for various applications.

\begin{figure*}[h]
    \centering 
     \includegraphics[width=\linewidth]{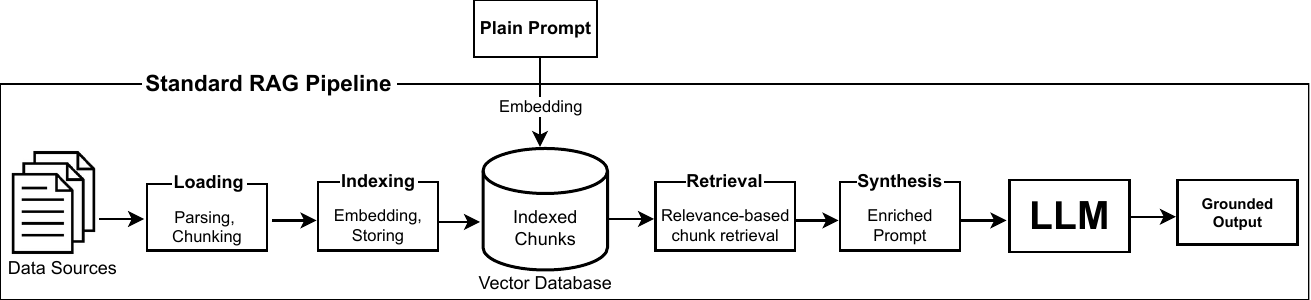}

    \caption{Standard pipeline of RAG System. It demonstrates an abstract view of the RAG pipeline, highlighting how a plain prompt can be enriched to generate a grounded output. Chunking, breaking down the data source into manageable segments, and embedding (converting these chunks into numerical representations or vectors) are the essential preprocessing steps that facilitate efficient storage, indexing, and retrieval of the chunks using a vector database.}
    \label{fig:typical_RAG_system_pipeline}
\end{figure*}

Fig. \ref{fig:typical_RAG_system_pipeline} demonstrates an abstract view of the standard RAG pipeline, highlighting how a plain prompt can be enriched to enable generating a grounded output. Chunking, breaking down the data source into manageable segments, and embedding (converting these chunks into numerical representations or vectors) are the essential preprocessing steps that facilitate efficient storage, indexing, and retrieval of the chunks using a vector database.

As illustrated in Fig. \ref{fig:typical_RAG_system_pipeline}, the standard RAG pipeline processes the end-user's text input (called prompt) as follows: It begins by embedding the prompt and searching through the indexed chunks in the vector database to find relevant chunks related to this prompt. If relevant chunks are found, the prompt is augmented with these chunks, forming an enriched prompt, which is then passed to the LLM. For the prompts that do not find relevant chunks, they are passed directly to LLM, missing the opportunity for augmentation. 

\begin{figure}[h]
    \centering 
    \includegraphics[width=\linewidth]{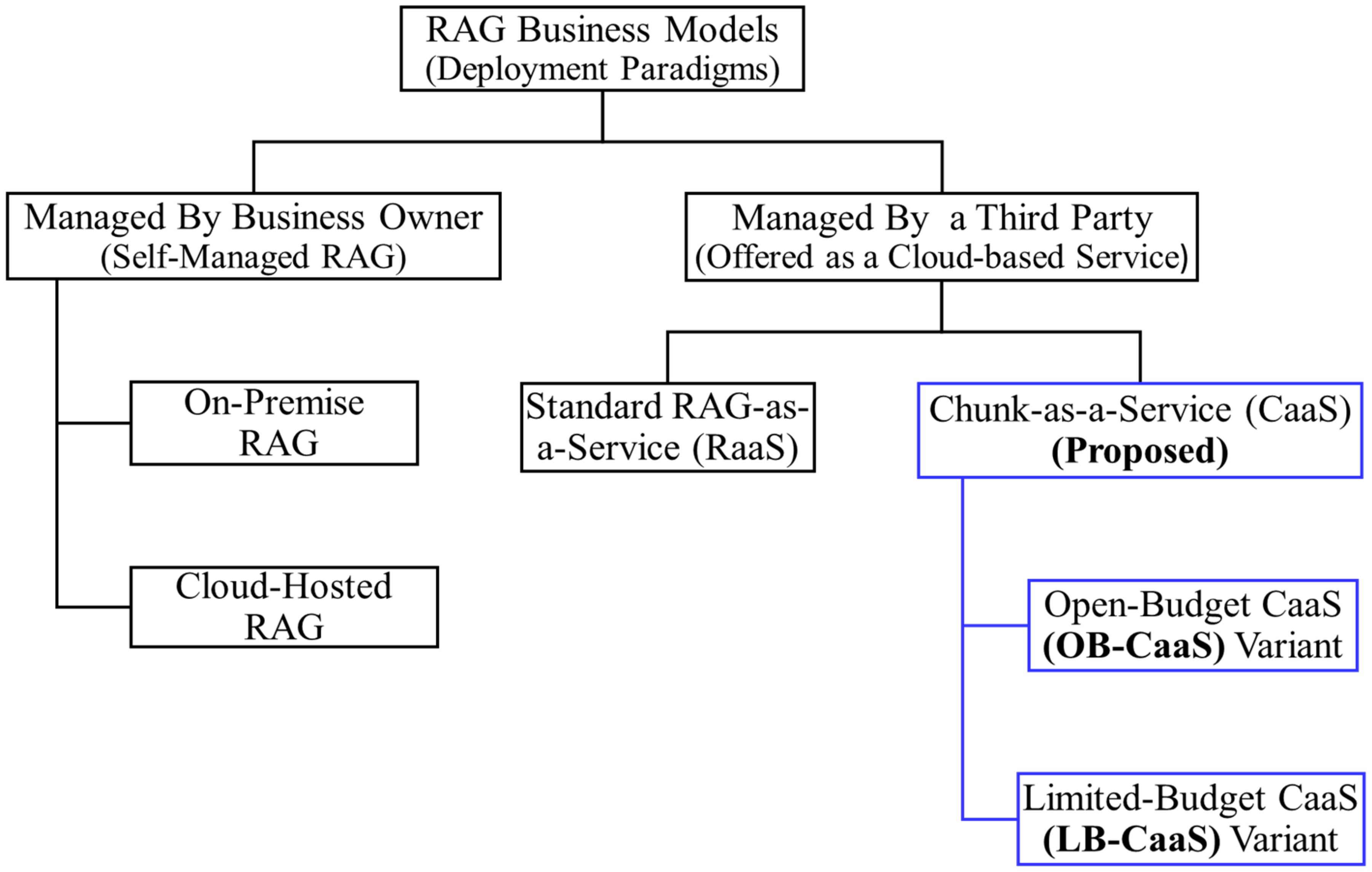}
    \caption{A high-level illustration highlighting the position of our proposed RAG variants, including OB-CaaS and LB-CaaS, among other related 
 RAG business models.}
    \label{fig:RAG_business_models}
\end{figure}

Standard RAG refers to the existing RAG variants with a vanilla pipeline, regardless of their deployment paradigm. This encompasses variants the business owner manages, including on-premise and cloud-hosted deployments. Standard RAG can also include variants deployed on the cloud and managed by a third party to be offered as a service, i.e., RAG-as-a-Service (RaaS), for interested users or organizations. The existing RAG variants are illustrated as black colored in Fig. \ref{fig:RAG_business_models}. 


In the standard RaaS, the cost is calculated based on the number of submitted prompts (RaaS API calls), regardless of whether the prompts find an opportunity to be enriched and no matter how the quality and quantity of the chunks used in the enrichment process. 
This indicates that standard RaaS is not cost-effective for limited-budget users who want to pay only for what they actually use. Even worse, RaaS is restrictive by design. It does not support enabling a chunk-based payment model \cite{guu2020retrieval, lewis2020retrieval}, even for the RaaS providers who are willing to consider such a payment approach. 

Depending on its implementation, RaaS can be accessed in various ways. One approach is through a single API that abstracts the complexity of managing the underlying required services, including embedding models, vector databases, and LLMs, along with their associated APIs (as in Vectara \cite{vectara_RAGaaS}). Another approach is through a RaaS API in addition to the APIs of the underlying services requested from the end user (as in LlamaIndex Cloud \cite{LlamaIndex_Cloud}). Regardless of the RaaS accessibility approach, RaaS providers consider the operational cost of managing the RAG pipeline as part of the established pricing. However, this operational cost does not necessarily consider the actually used chunks in the enrichment process, rendering the standard RaaS a vague and less transparent payment model that is potentially less cost-effective for the end users.

We address this limitation by proposing a variant that solely considers the actually used chunks in the prompt enrichment process. Specifically, we propose modifying the pipeline of the standard RaaS, particularly the indexing and retrieval stages, as follows.  
During the data source indexing stage and while chunking, a price is assigned to all indexed chunks.
This price considers various factors predetermined by the service provider and data source owner (see Section \ref{subsec:revenue_sharing}).
At the retrieval stage, all relevant chunks to a particular prompt are presented along with their associated price and relevance score. 
By incorporating the price into the chunks, end users can be easily billed primarily based on their actual utilization of chunks, along with other relevant operational costs, such as computing and storage. We term this chunk-based RaaS variant ``Chunk-as-a-Service (CaaS)'', highlighting that the utilized chunks are central to this model.
To the best of our knowledge, we are the first to introduce and coin the term CaaS.

Now, consider a scenario of a limited-budget user. 
The basic CaaS fails to support limited-budget end-users who cannot afford the surprise cost associated with utilized chunks, limiting the basic CaaS accessibility. Therefore, we propose a cost-aware CaaS variant called ``Limited-Budget CaaS (LB-CaaS).'' This variant addresses the shortcomings of basic CaaS, specifically its ineffectiveness in serving users with a limited budget.
In the cost-aware CaaS variant (LB-CaaS), the budget is predefined and allocated beforehand. Conversely, the basic CaaS variant does not require a predefined budget. Therefore, we refer to the basic CaaS as Open-Budget CaaS (OB-CaaS).
Fig. \ref{fig:RAG_business_models} illustrates, at a high level, the existing RAG variants, positioning our proposed OB-CaaS and LB-CaaS among them.
 
The selection of the chunks within the LB-CaaS variant is enabled by our proposed Utility-Cost Online Selection Algorithm (UCOSA). 
Considering budget constraints and the chunk's utility-cost tradeoff, UCOSA optimally selects a subset of plain prompts on the fly to enrich each prompt with a selected chunk, enabling cost-efficient generation of grounded output (detailed in Section \ref{lab:mathematical_modeling}).
UCOSA is a generic algorithm that can be applied in various scenarios using diverse representations of utility and cost. In this work, the utility is represented by the relevance score $R$ of the retrieved chunk, while the cost is represented by the associated price $P$ of each retrieved chunk.

The salient contributions of our work are as follows.
\begin{itemize}
    \item We introduce a novel RAG business model, Chunk-as-a-Service (CaaS), a chunk-based payment model that calculates usage costs based only on the chunks actually utilized in the prompt enrichment process. To the best of our knowledge, we are the first to introduce and coin the term CaaS.

    \item We propose Open-Budget CaaS (OB-CaaS), the basic variant of CaaS that makes standard RaaS transparent and cost-effective for end users.
    \item We propose Limited-Budget CaaS (LB-CaaS), a cost-aware CaaS that enhances OB-CaaS by making it more cost-effective and accessible for a broader range of end users.

    \item We propose the Utility-Cost Online Selection Algorithm (UCOSA) to enable the chunk selection process within LB-CaaS. UCOSA optimally selects chunks used for enriching prompts in an online manner. We provide a mathematical proof demonstrating the optimality of this algorithm.
    \item To validate UCOSA, we conduct extensive experiments demonstrating its effectiveness across various experimental scenarios and in comparison to multiple baselines.
\end{itemize}   

The rest of the paper is organized as follows. Section \ref{sec:motivation} situates our proposed RAG variants among the existing ones. In Section \ref{sec:proposed_RAG_variants}, we detail our proposed RAG variants, including OB-CaaS and LB-CaaS, and explain UCOSA, the enabler of LB-CaaS. We provide a mathematical formulation of our considered problem and solution in Section \ref{lab:mathematical_modeling}. We present the management of the proposed RAG variants in Section \ref{sec:management_of_the_proposed_RAG_variants}. Section \ref{sec:Experiments_and_results} illustrates the experimental results that demonstrate the efficacy of our proposed work. Lastly, Section \ref{sec:Conclusion_and_future_work} concludes our proposed work and highlights the potential future work.

\section{Motivation and Related Work} \label{sec:motivation}
This section highlights the motivations behind proposing CaaS variations, including OB-CaaS and LB-CaaS, and demonstrates a motivating example. Additionally, it presents the relevance of our proposed work to the existing literature.

\subsection{Motivations Behind Proposing OB-CaaS and LB-CaaS} \label{sec:motivations_related_works}

Compared to the standard RaaS, our proposed CaaS variants, including OB-CaaS and LB-CaaS, introduce several advantages that cannot be realized otherwise: 

\textbf{Empowering data source owners with a novel publishing paradigm:}
Traditionally, data source owners, such as book authors, have several options for gaining profit from their written works, including selling printed books, e-books, audiobooks, or all formats simultaneously. 
However, with advanced content-providing paradigms like RaaS, a publishing option that aligns with such paradigms still needs to be developed.
A RaaS provider cannot offer data source owners a chunk-based publishing paradigm that enables fair revenue-sharing, taking into account the value of the data source represented by chunks and their usage rate, due to the lack of enabling technology. Therefore, we introduced CaaS as an enabler of such a publishing paradigm.

\textbf{Granular profitability and quality content:} 
The proposed CaaS model allows proprietary data owners to sell their proprietary data with greater granularity (at the chunk level) compared to the traditional selling paradigms, where content is sold at a book or dataset level.

\begin{figure*}[]
    \centering 
    \includegraphics[width=0.9\linewidth]{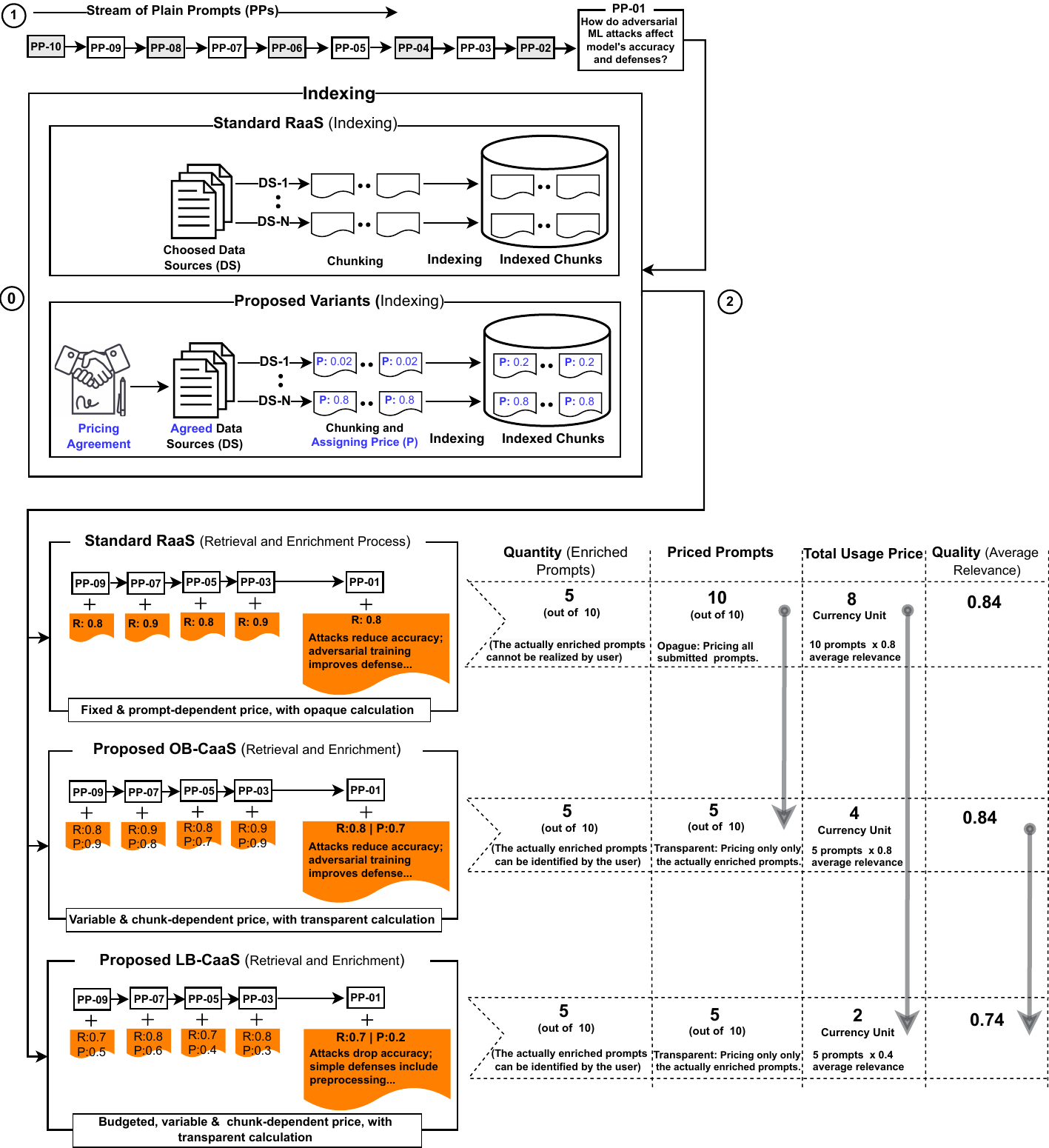}
    
    \caption{A toy example illustrating how various RAG variants, including Standard RaaS and our proposed OB-CaaS and LB-CaaS variants, handle a stream of prompts in terms of price calculation transparency, usage price, and quality of the generated output. This demonstration highlights that our proposed RAG variants enable a transparent price calculation. Additionally, it highlights that the total usage price is reduced by 50\% with OB-CaaS and 75\% with LB-CaaS. This price reduction occurs without compromising the quality of the generated outputs in OB-CaaS, and with only a 12\% decrease in quality when using LB-CaaS.
    }
    \label{fig:toy_illustrative_example}
\end{figure*}

CaaS enables this capability by adapting the typical pipeline to include pricing agreement during the indexing stage (details in \ref{fig:toy_illustrative_example}).
Data source owners, as well as CaaS providers, can earn profits based on the chunk's rate of usage and the predetermined prices of the chunks associated with their data source. This fair approach would encourage quality content creation due to the potentially established competition among content creators (e.g., authors) to maximize the use of their content by CaaS.

\textbf{Potential Applications:} Based on the above-mentioned motivations, there would be various potential applications, including the following:
\begin{itemize}
    \item Cloud-based RAG services: Cloud service providers can adopt and offer the CaaS model as a novel solution that provides flexible and cost-effective RAG service, charging users based on the actual data chunks they use rather than a flat fee.
    \item Online marketplaces for data: Platforms can be developed where data owners sell their content to the CaaS providers, ensuring they profit from every use of their sold content (at chunk level).
\end{itemize}

\subsection{Motivating Example}
\label{sec:Illustrative_example}

Fig. \ref{fig:toy_illustrative_example} demonstrates a toy example that compares various RAG variants, including Standard RaaS and our proposed OB-CaaS and LB-CaaS variants. The comparison considers handling a stream of prompts based on price calculation transparency, usage price, and quality of the generated output, represented by the average relevance scores of the chunks utilized for the enrichment process.

For clarity, we gray-colored the set of prompts expected not to be enriched by chunks---due to their irrelevance to the indexed chunks. For that, the enrichment process demonstrates a subset of the prompts (five out of ten). Additionally, for simplicity, we only textually illustrate the first prompt and the associated selected chunks. 

Additionally, Fig. \ref{fig:toy_illustrative_example} presents the indexing process (step 0), which is a prior step to prompting, aiming at efficiently storing data sources. The indexing compares the standard indexing with the proposed one, highlighting the differences, including the pricing agreement and assigning price. 
The pricing agreement aims at negotiating pricing between the data source owner and CaaS provider to determine the price per chunk (as detailed in Section \ref{subsec:revenue_sharing}).
Agreed data sources are chunked, assigned the agreed price per chunk per data source, and indexed. Differences in the compared indexing are highlighted in blue color.

Standard RaaS is opaque in terms of price calculation because pricing is prompt-dependent and not chunk-dependent. 
The actually enriched prompts cannot be realized by the end user. Users are charged based on the submitted prompts, regardless of whether they are enriched. Although only five prompts are enriched, the user is charged for all ten, resulting in a higher total usage price; double that of our proposed OB-CaaS. 
In RaaS, each prompt is priced uniformly (fixed price). To ensure a fair comparison of the enrichment process for the RaaS, we assigned the average price for the prompt enrichment in the OB-CaaS variant, which is 0.8 currency units. Consequently, the total price is 8 currency units (0.8 x 10 prompts).

Conversely, the proposed OB-CaaS and LB-CaaS are transparent in calculating the price associated with the prompts enrichment process
because pricing is chunk-dependent.
The transparent calculation is facilitated by assigning chunks with various prices, at the indexing stage, reflecting the value of their associated data source.
Thus, in these proposed variants, the number of priced prompts equals the number of enriched prompts, which is 50\% less compared to RaaS.

To realize the cost-effectiveness of utilizing the proposed OB-CaaS, we multiply the average cost of utilized chunks (0.8) by the number of the enriched prompts (5), resulting in 4 currency units, which is 50\% less than the associated price of RaaS without a sacrificing on the quality of the generated output. Quality is maintained at 0.84 for both RaaS and OB-CaaS.

Regarding LB-CaaS, since the proposed UCOSA strives to enrich prompts by striking a balance between the relevance score and the price of the selected chunks, the associated relevance scores and prices are less compared to OB-CaaS and RaaS. The average price for the utilized chunks is 0.4 currency units. Therefore, the total price of LB-CaaS is 2 currency units (0.4 x 5 prompts), which is 50\% less than OB-CaaS and 75\% less than RaaS, highlighting the higher budget utilization of LB-CaaS. This comes at the expense of only a 12\% reduction in the output quality.

For a valid comparison of the RAG variants, we can either use a fixed number of enriched prompts ($NEP$) and observe its impact on the total usage price and the quality of the enriched prompts, or we can allocate the same budget for all RAG variants and observe its impact on the quality and quantity ($NEP$) of the enriched prompts. 
For simplicity, and to avoid cluttering the demonstration with many prompts, we opted to assume a fixed $NEP$.

To illustrate LB-CaaS's budget efficiency, we allocated a limited budget of 2 currency units, whereas OB-CaaS and the standard RaaS operate without a predefined budget, highlighting their limitations in supporting budget-constrained scenarios.

In conclusion, end-user accessibility increases significantly when moving from RaaS to OB-CaaS, and even more so with LB-CaaS, making LB-CaaS the most feasible option for users with limited budgets.

\begin{table*}[b]
\caption{Related work summary. Compared to the standard RAG, our proposed CaaS variants excel in many aspects, including cost-effectiveness, user accessibility, and an enhanced pipeline that enables transparent usage cost calculation. In particular, LB-CaaS introduces a further enhancement to these aspects.}
\label{fig:related_work_summary}
\centering
\resizebox{0.999\textwidth}{!}{%
\begin{tabular}{|l|l|l|l|l|l|l|l|l|l|}
\hline
\textbf{Ref.} & \multicolumn{2}{l|}{\textbf{RAG Business Model}} & \textbf{\begin{tabular}[c]{@{}l@{}}Cost\\ Effectiveness\end{tabular}} & \textbf{\begin{tabular}[c]{@{}l@{}}Enhanced\\ Pipeline\end{tabular}} & \textbf{\begin{tabular}[c]{@{}l@{}}Retrieved Chunks\\ Characteristics\end{tabular}} & \textbf{\begin{tabular}[c]{@{}l@{}}Chunk Selection\\ Criteria\end{tabular}} & \textbf{\begin{tabular}[c]{@{}l@{}}End User\\  Accessibility\end{tabular}} & \textbf{\begin{tabular}[c]{@{}l@{}}RAG Service\\ Type\end{tabular}} & \textbf{\begin{tabular}[c]{@{}l@{}}Payment Model\end{tabular}} \\ \hline
\multirow{2}{*}{\begin{tabular}[c]{@{}l@{}}
\cite{LlamaIndex_Cloud, vectara_RAGaaS} \\
\cite{nuclia_RaaS,geniusee_RaaS} \\
\cite{amazon_bedrock_RaaS} \\
\end{tabular}}
& \multirow{2}{*} & On-Premise RAG & Low & No & Relevance Score & Relevance Score & N/A & N/A & \begin{tabular}[c]{@{}l@{}}Paying for in-house \\ expertise and\\ hardware  resources\end{tabular} \\ \cline{3-10}

& \multirow{3}{*}
{Standard RAG}  & Cloud-Hosted RAG & Low & No & Relevance Score & Relevance Score & N/A & N/A & \begin{tabular}[c]{@{}l@{}}Paying for cloud \\ resources and\\management staff\end{tabular} \\ \cline{3-10}

 &  & RAG-as-a-Service (RaaS) & Low & No & Relevance Score & Relevance Score & Low & RaaS & \begin{tabular}[c]{@{}l@{}}Pay-as-you-Go\\ (Open Budget)\end{tabular} \\ \hline
\multirow{2}{*}{\textbf{This Work}} & \multirow{2}{*}{\textbf{Proposed RAG}} & \textbf{OB-CaaS} & \textbf{Medium} & \textbf{Yes} & \begin{tabular}[c]{@{}l@{}}Relevance Score \\ and Chunk Price\end{tabular} & Relevance Score & \textbf{Medium} & \textbf{CaaS} & \begin{tabular}[c]{@{}l@{}}Pay-as-you-Go\\ (Open Budget)\end{tabular} \\ \cline{3-10} 
 &  & \textbf{LB-CaaS} & \textbf{High} & \textbf{Yes} & \begin{tabular}[c]{@{}l@{}}Relevance Score \\ and Chunk Price\end{tabular} &\begin{tabular}[c]{@{}l@{}}Relevance Score, \\ Chunk's Price, and \\ Remaining Budget\end{tabular} & \textbf{High} & \textbf{CaaS} & \textbf{Limited Budget} \\ \hline
\end{tabular}
}
\end{table*}
\subsection{Related Work}
\label{sec:related_work}

LLMs have transformed numerous industries through their exceptional capabilities. However, they still have several limitations \cite{brown2020language}, including restrictions on context length \cite{xuretrieval}, issues with outdated knowledge \cite{roberts2020much}, hallucinating untrue output \cite{ji2023survey}, and challenges with proper attribution \cite{menick2022teaching}. Each of these limitations is the focus of active research aimed at addressing them. For example, hallucinations are being tackled through augmentation with external databases (RAG), and hallucination detection methods \cite{kuhn2023semantic, farquhar2024detecting, hron2024training}. 
RAG \cite{guu2020retrieval, lewis2020retrieval} has been explored in various variants in the literature.
Gao et al. \cite{gao2023retrieval}
demonstrated three variants of RAG: naive RAG, advanced RAG, and modular RAG. Both naive and advanced RAGs consist of pipeline stages, including indexing, retrieval, and generation, following a sequential process. However, unlike naive and advanced RAGs, which maintain a sequential flow, the modular RAG variant incorporates iterative and adaptive retrieval strategies. 

To differentiate various RAG systems, their limitations should be identified, and their performance should be evaluated. To this end, Barnett et al. \cite{barnett2024seven} report on the failure points of RAG systems, emphasizing that validation of an RAG system is only truly feasible during its operation and the robustness of an RAG system is developed over time rather than being fully established at the beginning. Additionally, Chen et al. \cite{chen2024benchmarking} introduce the Retrieval-Augmented Generation Benchmark (RGB) to identify the limitations and evaluate the effectiveness of RAGs when they integrate with different LLMs.

Standard RAG can be implemented on-premise, cloud-hosted, or offered as a cloud service, i.e., RAG-as-a-Service (RaaS) \cite{LlamaIndex_Cloud, vectara_RAGaaS,nuclia_RaaS,geniusee_RaaS,amazon_bedrock_RaaS} (as illustrated in Fig. \ref{fig:RAG_business_models}). However, these existing implementations are not cost-effective: managing on-premise RAG requires investment in infrastructure and expensive expertise. Similarly, cloud-hosted requires payment for cloud resources and management staff. Additionally, both on-premise and cloud-hosted RAGs require laborious updating of the indexed data sources. On the other hand, RaaS is more cost-effective compared to the on-premise and cloud-hosted implementations. However, the end user keeps paying for all the submitted prompts, regardless of the quality and quantity of the actually utilized chunks.

In addition, in standard RAG implementations, the retrieved chunks are selected solely based on their relevance score, without incorporating a data source-dependent cost to better define and differentiate the chunks. This limitation in the granular description of the chunks makes identifying the utilized chunks challenging. Consequently, the standard RaaS implementation offers less flexibility and is provided without considering the actual usage of the chunks. The lack of granularity in calculating the utilized chunks makes the operational cost higher, which, in turn, impacts the overall service cost, resulting in an expensive pricing model. Therefore, RaaS accessibility is restricted to end users who can afford such costs. This highlights a clear gap in the existing business models of standard RAG, including RaaS.

To bridge this gap, we propose modifying the standard RAG pipeline to incorporate a weight (a price) to the chunks of each data source based on predetermined factors set by the data source owner and the service provider (more details in Section \ref{subsec:revenue_sharing}). 
Therefore, the utilized chunks can be quantified, facilitating the introduction of our proposed CaaS.
Table \ref{fig:related_work_summary} summarizes the comparison between standard RAG and our proposed CaaS variants, including OB-CaaS and LB-CaaS.
Compared to the standard RaaS, our proposed CaaS variants excel in many aspects, including cost-effectiveness, user accessibility, and an enhanced pipeline that enables transparent usage cost calculation. Notably, LB-CaaS introduces a further enhancement to these aspects.

With the ongoing competition among RaaS providers, their pricing plans continue to evolve \cite{vectara_price_plan, llamaIndex_price_plan, nuclia_price_plan}. Regardless of the offered prices, adopting our proposed CaaS variants is supposed to achieve further cost reductions as it optimizes the underlying process itself: the pipeline of the standard RAGs, including RaaS.

\textbf{Learning-Based Online Selection Techniques:} 
There are learning-based online selection techniques, such as RL- or ML-based approaches, that are used for online selection \cite{tang2025mba, belgacem2022dynamic, ghasemi2018cost}. However, these techniques are complex. They are associated with pre-training overhead. For example, Tang et al. \cite{tang2025mba} proposed an RL-based framework to select a retrieval method based on the complexity of the served prompt, suggesting that each prompt requires a different retrieval approach. 
In contrast, our proposed UCOSA is a lightweight and training-free algorithm that works on the fly without training overhead. Such a simple and lightweight technique ensures the practicality and simplicity of our proposed budget-constrained variant (LB-CaaS).

\textbf{Budget-Constrained Cloud Resource Allocation:}
Inspired by the established cloud computing services, such as software as a service, platform as a service, and storage as a service \cite{younis2024comprehensive}, where cloud resources are abstracted and offered as independent services to facilitate utilization and integration with other systems and services in a cost-effective and efficient manner, we propose offering the key building block in RAG systems ``the chunk'' as a service to gain the same merits of the ``as-a-service'' paradigm, including the smooth integrations with other systems in a cost-effective and efficient way.\newline
In particular, our proposed budget-constrained variant (LB-CaaS) is related to the well-established principles of budget-constrained optimization in cloud resource allocation \cite{belgacem2022dynamic, ghasemi2018cost} as follows. Both approaches aim to optimize utility under cost constraints: cloud resource allocation aims to maximize the utilization of computational resources, while LB-CaaS aims to cost-effectively utilize information resources (chunks), considering their relevance to the served prompt and their associated price. 
However, they differ in the type of resources and the decision-making paradigm. Cloud resource allocation optimizes computational resources (CPU, memory, storage, bandwidth)\cite{belgacem2022dynamic, ghasemi2018cost, kumar2020autonomic} and typically uses prediction-based provisioning \cite{belgacem2022dynamic, ghasemi2018cost} where future resource requests are predicted using machine learning models. In contrast, our LB-CaaS model optimizes the information resources (chunks) selection, utilizing the proposed online selection algorithm (UCOSA), where each chunk must be immediately accepted or rejected upon arrival without knowledge of future chunks.

\section{Chunk-as-a-Service (CaaS): Proposed RAG Variants}
\label{sec:proposed_RAG_variants}
This section details our proposed work. First, we introduce the Open-Budget CaaS (OB-CaaS), the basic variant. Then, we present the Limited-Budget CaaS (LB-CaaS), the cost-aware variant. 

\subsection{Proposed OB-CaaS Variant}
\label{sec:proposed_OB-CaaS}

Standard RaaS lacks the capability enabling quantifying the actually utilized chunks in the prompt's enrichment process.
For example, data source owners, such as book authors, are unable to align their sales with RaaS's operational style. Ideally, they would earn profits reflecting the rate of usage and the value of the utilized chunks associated with their indexed data. This would be a fair approach and would encourage creating high-quality data source content due to the expected competition among content creators to produce valuable and competitive content to find its opportunity to be selected and utilized in the prompt enrichment process.
Therefore, we propose the 'Chunk-as-a-Service' (CaaS) business model to address the limitations of standard RaaS, including the lack of chunk-based usage price calculation. Similar to standard RaaS, our proposed CaaS is intended to be deployed on the cloud and managed by a third party. However, it offers a chunk-based, transparent usage price calculation, and more cost-effective solution.

CaaS adapts the data source indexing process to assign prices to the indexed chunks based on predetermined criteria set by the CaaS provider and the data source owner (detailed in Section \ref{sec:Illustrative_example}). In its basic variation, CaaS is considered an Open-Budget CaaS (OB-CaaS) model. It does not require the end users to set a predefined budget. The usage price is calculated based on the associated prices of the utilized chunks, which are selected considering one factor: their degree of relevance to the prompts regardless of the associated prices.
The key parties involved in the CaaS model are the CaaS provider, the data source owner, and the CaaS subscriber (the end-user).
The data is provided by the data source owner, managed by the CaaS provider, and utilized by the CaaS subscriber.

\begin{figure*}[t]
    \centering 
    \includegraphics[width=\linewidth]{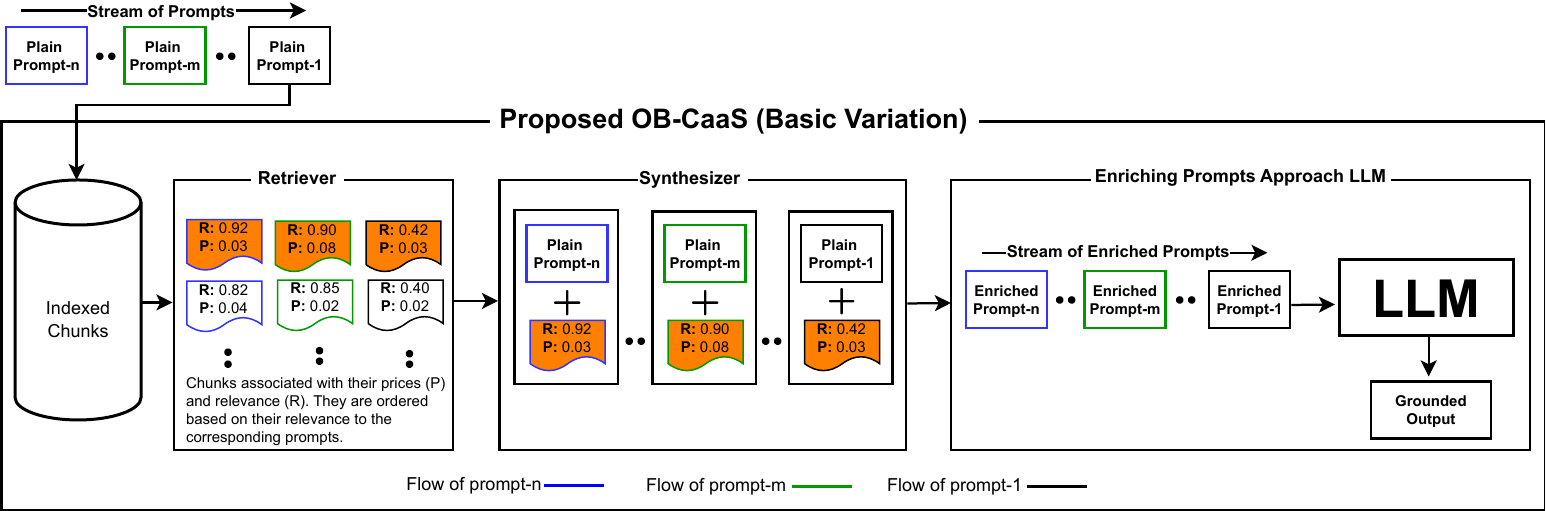}
    \caption{\textit{The pipeline of the OB-CaaS}. 
    Unlike standard RaaS where the indexed chunks are not incorporated with a price reflecting the value of their data sources, the indexed chunks in the proposed OB-CaaS have a negotiation-based price set during the price agreement stage between the data source owner and the service provider (detailed in Section \ref{sec:Illustrative_example}). This incorporation of the prices into the chunks enables quantifying the end user's utilized chunks and facilitates the introduction of the CaaS model in its basic variant (OB-CaaS).
}
    \label{fig:proposed_RAG_basic_variation}
\end{figure*}

Fig. \ref{fig:proposed_RAG_basic_variation} illustrates the pipeline of our proposed OB-CaaS. It highlights our contributions in the indexing and retrieval stages. Compared to the standard RAG pipeline, where the price of usage does necessarily not consider the actual utilization of chunks, the pipeline of our proposed OB-CaaS considers the price of the utilized chunks. This is due to the predetermined prices assigned to the chunks of the indexed data sources. The cost is simply the total price associated with the utilized chunks.
Different colors are used to visually distinguish the flow of individual prompts (Prompt-1, Prompt-m, Prompt-n) through the CaaS pipeline.
Although OB-CaaS surpasses standard RaaS in many aspects, it still shares a key limitation: it is less accessible to a broader range of end users, particularly those with a limited budget. This is because OB-CaaS lacks an intelligent chunk selection functionality that prioritizes chunk selection based on both relevance score and price, preventing it from being a cost-effective option for limited-budget users. Therefore, we propose a cost-aware variant beyond the OB-CaaS, as explained below (Section \ref{sec:proposed_LB-CaaS}).

\subsection{Proposed LB-CaaS Variant}
\label{sec:proposed_LB-CaaS}

LB-CaaS addresses the shortcomings of OB-CaaS, specifically the relative cost-ineffectiveness and lack of accessibility for limited-budget users. This variant enhances the pipeline of OB-CaaS pipeline as follows:

\textbf{At ``Retriever'' stage}, for each plain prompt in the stream, instead of the typical process of retrieving indexed chunks based on the relevance score and then selecting the most relevant one, the relevant chunks are only retrieved\footnote{The retrieved chunks can be stored in a cache, which is emptied after the selection of a chunk to prepare for the next prompt.}, while the selection process is delegated to a new contributed stage named the ``Chunk's Selection Process'' to optimize the selection considering the remaining budget, chunk's relevance score, and chunk's price.

\textbf{At the ``Chunk's Selection Process'' stage},
UCOSA selects chunks with relevance-to-price ratios above a budget-dependent threshold, forming a set of candidate chunks (CCH). From within a CCH, the chunk with the highest relevance score is then selected (highlighted in orange) and passed to another stage named “Enriching prompts with the selected
chunks
\textbf{Stage ``Enriching prompts with the selected chunks (Prompt Online Selection)''} 
begins by checking the availability of a selected chunk associated with each prompt. 

If a chunk is selected by the previous stage, the associated plain prompt utilizes the RAG’s synthesizer component to compose an enriched prompt and pass it to the LLM to get a grounded output. However, if none of the chunks were selected, the corresponding plain prompt is switched to the LLM directly, generating traditional output.

Selecting a chunk in this context means that the associated prompt is chosen in an online fashion for the enrichment process, and this decision is irrevocable. 

\begin{figure*}[!t]
    \centering 
    \includegraphics[width=\linewidth]{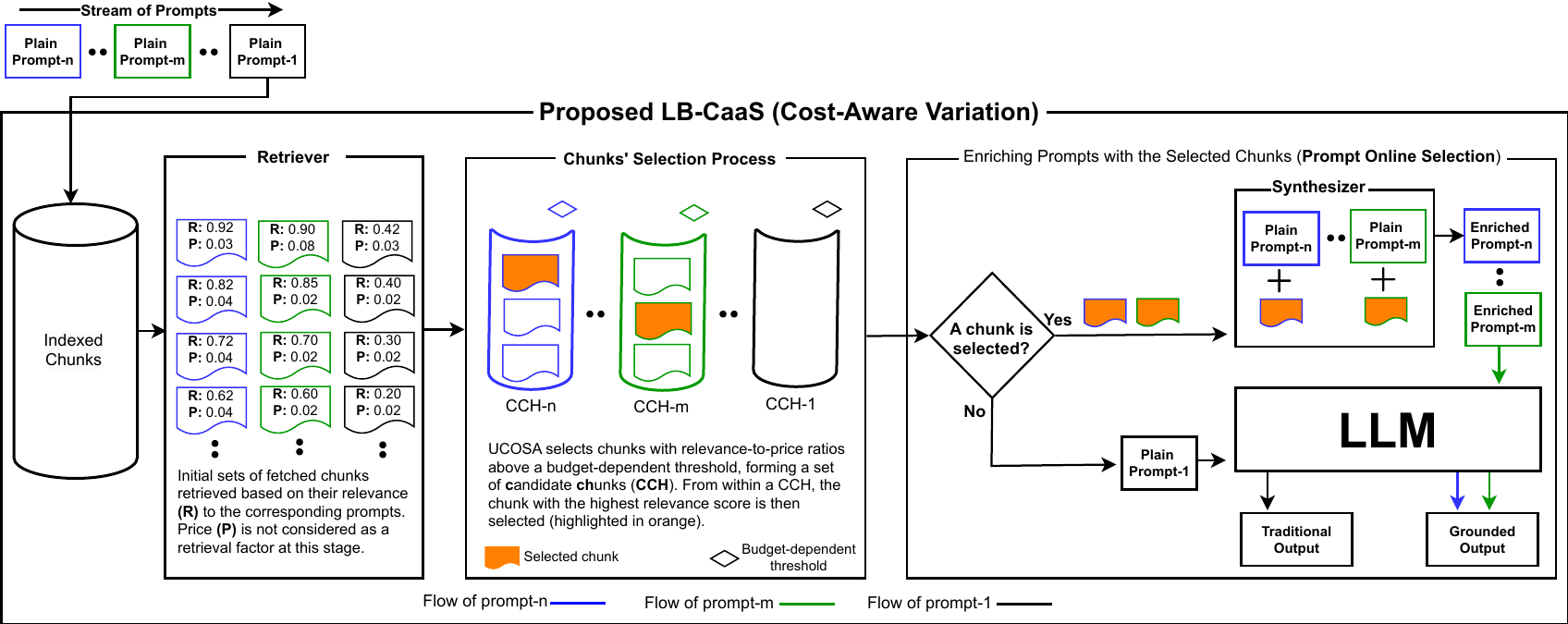}
    \caption{\textit{The pipeline of the proposed LB-CaaS}. A cost-aware variant that is more cost-effective than the OB-CaaS, allowing accessibility for end users with limited budgets. Utilizing UCOSA, a proposed online selection algorithm, a subset of plain prompts is selected optimally on the fly for the enrichment process, enabling cost-efficient generation of grounded output.
}
    \label{fig:proposed_RAG_enhanced_variation}
\end{figure*}

The algorithmic representation of Figure \ref{fig:proposed_RAG_enhanced_variation} is illustrated in line with the presentation of the UCOSA (see Algorithm \ref{alg:LB_CaaS_Rep_Algorithim}), the enabler of LB-CaaS. The full details of UCOSA are provided below in Section \ref{lab:mathematical_modeling}.

\section{Online Selection Algorithm: Modeling and Analysis} \label{lab:mathematical_modeling}

This section details the mathematical modeling of our proposed online selection algorithm (UCOSA), including the settings, optimization technique, algorithmic representation, and competitive ratio analysis

\subsection{Settings}\label{subsec:Settings}
There are $\mathcal{I} = \{1, 2, \ldots, i, \ldots, I\}$ plain prompts and $\mathcal{J} = \{1, 2, \ldots, j, \ldots, J\}$ chunks. Enriching the $i$-th prompt with the $j$-th chunk achieves a relevance score of $R_{ij}$ with a price of $P_{ij}$\footnote{The price of a chunk may be the same, regardless of the prompts. However, we assume a different price for different prompts as a generalization.}. The users (\emph{i.e.,} the providers of the prompts) have a budget $B$. The objective is to find an assignment of prompts to chunks that maximizes the overall Relevance score, without exceeding the budget.

\subsection{Optimization Technique} \label{sec:optimization_technique}
The problem of chunk selection is formulated as an optimization problem. Let:

$x_{ij}  =\left\{ \begin{array}{rl}
	1 &\mbox{if the $i$-th prompt is enriched}\\
	&\mbox{with the $j$-th chunk}\\
	0 &\mbox{otherwise.}
\end{array}\right.$

Then the optimization becomes:

\begin{equation}
	\max \sum_i \sum_j x_{ij} R_{ij}
\end{equation}  
Subject to:

\begin{align}
	&\sum_i \sum_j x_{ij} P_{ij} \leq B \label{const1}\\
	& \sum_j x_{ij} \leq 1 \quad \forall i \label{const2}
\end{align}

Where Constraint~\eqref{const1} states the budget limit, and Constraint~\eqref{const2} states that a prompt can be enriched with 1 chunk at most. The optimization technique assumes that all information, regarding prompts and chunks, are known in advance. The formulation above is for LB\_CaaS. Note that setting $B \rightarrow \infty$ converts it to OB\_CaaS.

\subsection{Online Selection Algorithm (UCOSA)}
Under the online settings, the prompts are fed to the LB\_CaaS RAG as a stream (\emph{i.e.,} one by one). For each prompt, an initial set of chunks are retrieved (by the Retrieval stage) based on their Relevance score, $R_{ij}$. At the Chunk Selection stage, a decision must be made to choose, at most, one chunk to enrich the prompt, considering the Relevance score, the price $P_{ij}$, and the available budget $B$. Note that $R_{ij}$ and $P_{ij}$ are only known when the prompt is fed to the RAG system, not before. A decision for a prompt must be made before considering the next prompt, and these decisions cannot be revoked afterward. To meet these requirements, we propose the Utility-Cost Online Selection Algorithm (UCOSA).
With respect to UCOSA, the relevance is required to be represented as a scalar. However, the way this scalar is obtained is completely under the control of the service provider. It could be a simple relevance measure, or could be obtained through a weighted function of several variables (e.g., diversity, novelty).
Relevance is a well-established metric used in the literature to calculate the similarity between two texts and is often used interchangeably with the term similarity  \cite{lin2025scorerag,wang2024rear}.

UCOSA is a generic algorithm that can be applied in various scenarios using diverse representations of utility and cost. In this work, the utility is represented by the relevance score $R$ of the retrieved chunk, while the cost is represented by the associated price $P$ of each retrieved chunk.
UCOSA is a specialized online selection variant suited to the settings of our proposed LB-CaaS RAG, where chunks are defined by relevance and price.
It requires assigning a price at the time of selection (assigning prompts the chunks), while the assignment of prices to the chunks is part of the indexing stage.

Before we state the details of UCOSA, some assumptions must be stated:
\begin{itemize}
	\item There exists non-trivial bounds on the values of $\frac{R_{ij}}{P_{ij}}$. That is, $0 < L \leq \frac{R_{ij}}{P_{ij}} \leq U < \infty$.
	\item $\frac{P_{ij}}{B} \leq \epsilon$, where $\epsilon$ is close to 0. 
\end{itemize}
The first assumption means that neither the Relevance score can be infinite, nor the price can be 0. Otherwise, the solution becomes trivial (by selecting the chunk with the infinite score, or selecting the chunk with a zero price). The second assumption states that the price of a chunk is relatively small compared to the available budget.
This assumption aligns well with real-world cloud service scenarios, where a predefined budget typically serves multiple requests (storage, computation, or retrieval). Such a realistic assumption helps define an appropriate business model for both service providers and end users. In addition, users usually tend to get services from providers that offer a small price compared to their budget. This low-pricing model is common and widely adopted by cloud service providers, including AWS Lambda \cite{AWS2025}.

UCOSA selects chunks with relevance-to-price ratios above a budget-dependent threshold, forming a set of candidate chunks (CCH). From within a CCH, the chunk with the highest relevance score is then selected.

Mathematically speaking, let $z_i$ be the fraction of budget used when the $i$-th prompt is fed to LB\_CaaS, and define $\Psi(z_i) = \big({Ue/L}\big)^{z_i} (L/e)$, where, $e$ is the natural number. Let $Q_i$ be the set of chunks that satisfy $\frac{R_{ij}}{P_{ij}} \geq \Psi(z_i)$. If $Q_i$ is empty, the $i$-th prompt is passed to LLM without enrichment. Else, we pick the chunk from $Q_i$ that gives the maximum Relevance score, and enrich the prompt with that chunk.
So, the impact of UCOSA on threshold $\Psi(z_i)$ (the budget-dependent threshold) is as follows: the more UCOSA selects chunks and uses a higher fraction of the budget, the threshold becomes higher, and UCOSA becomes stricter in selecting the next chunks.
UCOSA is shown in Algorithm~\ref{alg:LB_CaaS_Rep_Algorithim}.
\begin{algorithm}
        \caption{UCOSA} 
	\label{alg:LB_CaaS_Rep_Algorithim}
	\textbf{Inputs:}
	\begin{itemize}
		\item \textbf{Plain\_Prompts:} Stream of plain prompts
		\item \textbf{B:} Allocated Budget
	\end{itemize}
	
	\textbf{Outputs:}
	\begin{itemize}
		\item \textbf{Selected\_Chunks:} A subset of the retrieved chunks 
		\item \textbf{Enriched\_Prompts:} A subset of Plain\_Prompts enriched with their associated selected chunks. 
	\end{itemize}
	
	\begin{algorithmic}[1]
		\STATE{Initially, $z_i = 0$}
		\FOR{Each prompt $i$ in \textbf{Plain\_Prompts}} 
		\STATE{\textbf{Retrieval Stage:}}
		\STATE{fetch the relevant chunks $\mathcal{J}_i$}
		\STATE{\textbf{Selection Stage:}}
		\STATE{Compute $z_i, \Psi(z_i)$}
		\STATE{Set $Q_i = \{j \in \mathcal{J}_i \vert \frac{R_{ij}}{P_{ij}} \geq \Psi(z_i)\}$}
		\STATE{find $\hat{j} = argmax_{j \in Q_i} R_{ij}$}
		\STATE{\textbf{Enrichment Stage:}}
		\IF{No chunk is selected}
		\STATE{Pass the prompt to LLM without enrichment}
		\ELSE
		\STATE{Enrich prompt $i$ with chunk $\hat{j}$}
		\ENDIF
		\ENDFOR
		
	\end{algorithmic}
\end{algorithm}
 Appendix A demonstrates that the performance of UCOSA is competitive with respect to the optimization formulation.

\section{Management of the Proposed CaaS Model}
\label{sec:management_of_the_proposed_RAG_variants}

\subsection{Indexing and Retrieval Management} \label{subsec:indexing_management}

RAG providers facilitate building and managing RAG pipelines. For example, frameworks such as Haystack \cite{Haystack} can be used to flexibly build custom RAG pipelines for on-premise use or cloud deployment.
A RAG pipeline comprises key components, including indexing and retrieval. Typically, RAG providers offer options for managing these components either by using a built-in vector database, such as the Vector Store Index \cite{LlamaIndex_Vector_Store} of LlamaIndex, or by utilizing third-party vector database engines, such as FAISS \cite{META_FAISS}, Pinecone \cite{Pinecone_Vector_Database}, Chroma \cite{Chroma_Vector_Database}, and Elasticsearch \cite{Elastic_Search_Relevance_Engine}.

For several reasons, for our proposed RAG variants, including OB-CaaS and LB-CaaS, managing the indexing stage requires a built-in vector database. First, once the data source owner and CaaS provider sign a revenue-sharing agreement (discussed in Section \ref{subsec:revenue_sharing}), the loaded data becomes a proprietary asset for both parties, necessitating its protection through internal management and storage. Second, the built-in vector database allows us to adapt the pipeline to support the proposed RAG variants. For instance, the indexing stage can be adapted to assign chunk prices during the chunking and storing steps. The retrieval stage can also be modified to incorporate budget-constrained online selection using UCOSA. This means that utilizing an external vector database (e.g., through an API) is not a viable option.

\subsection{Revenue Sharing Management}
\label{subsec:revenue_sharing}

Various subjective factors determine the reputation and reliability of the service provider of the proposed CaaS variants and the value of the data source intended to be loaded and indexed within these proposed CaaS variants. Therefore, we adopt a negotiation-based agreement for the revenue-sharing of our proposed CaaS variants. To ensure transparent and reliable revenue-sharing, a relevant well-established system, such as blockchain-based smart contracts \cite{OpenZeppelin_SmartContract, Consensys_SmartContract} or a dedicated revenue management platform \cite{Zuora_Revenue_Sharing, Chargebee_Revenue_Sharing}, is assumed to be integrated with the proposed CaaS payment model to handle the revenue-sharing process. These systems can provide automated, immutable, and transparent tracking of revenue splits, ensuring both parties have confidence in the fairness and accuracy of the agreement.

We assume that CaaS providers make rational agreements that reflect the true value of the negotiated data source. Otherwise, users might end up paying more for lower-quality content. 

\section{Experiments and Results}
\label{sec:Experiments_and_results}
In this section, we conduct extensive experiments to empirically investigate the performance of UCOSA.

\subsection{Performance Evaluation Metrics and Baselines}

\textbf{Evaluation Metrics:} For our proposed LB-CaaS, the impact on the quantity and the quality of the generated output are the two key factors that should be considered during the evaluation process. 
 
\begin{itemize}
    \item Quantity: How many enriched prompts does the proposed RAG provide (the number of the online selected prompts)?
    \item Quality: How impactful these enriched prompts are on the generated output?
\end{itemize}

We use the chunk's Average Relevance ($AR$) as a proxy to measure the quality of the generated output. The more relevant the chunk is to the plain prompt, the higher the potential that the associated enriched prompt incorporates to generate quality (reliable) output. 
For quantity, we count the Number of Enriched Prompts ($NEP$): the set of online selected prompts augmented by the selected chunks.
In this work, we use the terms quality and relevance score interchangeably.

In limited-budget selection scenarios, considering quality ($AR$) or quantity ($NEP$) alone does not guarantee optimal performance. High-quality output may sometimes coincide with a significant decrease in $NEP$. Conversely, scenarios with high $NEP$ might exhibit low quality (i.e., low $AR$). Therefore, we propose a performance evaluation metric that incorporates both quantity and quality to comprehensively assess the impact of our proposed approach on the generated outputs.
$NEP$ denotes quantity and $AR$ denotes quality.
Given that $AR$ is a value greater than 0.0 and less than or equal to 1.0, where values closer to $0.0$ are considered least relevant and those closer to $1.0$ are highly relevant, multiplying $NEP$ by quality $AR$ serves as a fair performance evaluation metric. This metric ($NEP \times AR $) penalizes low-quality outputs and rewards those with higher quality, providing a balanced assessment.

Table \ref{table:annotations_summay} demonstrates the descriptions of the notations used in the conducted experiments.

\begin{table}[h!]
\caption{Description of the notations used in the conducted experiments.}
\centering
\begin{tabular}{|c|p{6 cm}|}
\hline
\textbf{Annotation} & \textbf{Description} \\
\hline
$NEP$ & Number of Enriched Prompts. The higher the better. \\
\hline
$AR$ & The Average Relevance score of the selected chunks. The higher the better. \\
\hline
($NEP \times AR$) & This metric rewards high relevance and penalizes low relevance, with higher scores being better.\\
\hline
B & The allocated budget \\
\hline
P & Price per chunk per data source \\
\hline
Total Relevance & The total relevance scores of the enriched prompts.
\\
\hline
\end{tabular}
\label{table:annotations_summay}
\end{table}


\textbf{Baselines:} We use offline selection and ``relevance-greedy'' selection algorithms as baselines to compare the performance of our proposed approach. In online selection, prompts are chosen on the fly, and the decision cannot be revoked. Offline selection represents the ideal selection scenario, where all prompts are available for analysis beforehand. In contrast, ``relevance-greedy'' selection represents the typical selection approach that is considered by standard RAG, including the RaaS variant.

\subsection{Experimental Setup}
We used LlamaIndex RAG framework \cite{llamaindex_2024} to conduct our experiments. Within LlamaIndex, we utilized ChatGPT-3.5 \cite{openai_chatgpt3.5} as LLM
and OpenAI text-embedding-ada-002 \cite{LlamaIndex_Embedding} as an embedding model.

Our extensive experiments run on top of the modified RAG's pipeline that we propose to accommodate our proposed CaaS variations. The typical stages of RAG's pipeline include data source loading, indexing, retrieval, synthesis, and generation. Our proposed work mainly contributes to the indexing and retrieval stages.

For indexing, we used LlamaIndex default vector database: Vector Store Index \cite{llamaindex_2024}. 
The indexed chunks are then stored in local storage using LlamaIndex Storage Context to avoid the overhead of re-indexing.
Our proposed UCOSA is agnostic to the choice of similarity metric. We used cosine similarity as a representative metric for its popularity in RAG-related works \cite{sarthi2024raptor,li2024retrieval}. It is the de facto standard and the default setting in many retrieval systems and RAG frameworks, including LlamaIndex \cite{LlamaIndex2025_similarity_metrics, LlamaIndex_Cosine_Similarity}, where it is used to compute the similarity between two embedding vectors: one represents the prompt and the other represents an indexed chunk. Moreover, cosine similarity excels in capturing the semantic relationship between embeddings within high-dimensional vector spaces \cite{cosine_similarity_effectiveness}.
We set the parameter ``similarity-top-k'', the number of retrieved chunks, to 20, giving a sufficient chance to our proposed UCOSA to select chunks while striking a balance between the relevance score (similarity) and the price.

Chunk size influences UCOSA’s performance because it determines the quality of the candidate chunks available for selection. With the optimal chunk size, UCOSA’s selections maximize the quality of the generated output, whereas smaller or larger chunks make UCOSA less effective, though it remains functional. We set the chunk size to 1024 tokens because previous evaluation studies, conducted by leading RAG framework providers, including LlamaIndex \cite{LlamaIndex2025_Chunksize_Evaluation}, found it optimal. They have already investigated the impact of chunk size on relevance and found that 1024 is the ideal chunk size for the best relevance score because this size is large enough to capture context while being manageable for processing.
Additionally, we set chunk overlap to 20 tokens as it helps ensure smoother transitions between chunks and avoids information loss at chunk boundaries.

We designed the experiments to handle prompts with content from fields related to the areas of the indexed data sources. This design allows the retrieval stage to find candidate chunks that are relevant to the prompt. Thus, if UCOSA finds one of the candidate chunks feasible to be selected, the plain prompt is augmented with this chunk, composing an enriched prompt (online prompt selection). 

For this purpose, we loaded and indexed a few data sources (e-books) that cover the areas of adversarial robustness and selection algorithms, namely the books, Adversarial Robustness for Machine Learning \cite{chen2022adversarial}, Strengthening Deep Neural Networks \cite{warr2019strengthening}, and Algorithms to Live By: The Computer Science of Human Decisions \cite{christian2016algorithms}. We assigned predetermined prices of 0.8, 0.2, and 0.1 currency units for the chunks of each book, respectively, introducing price variability to facilitate the algorithm's prioritization during selection. 
We assume that the predetermined price for chunks of each book is set following a negotiation-based agreement between e-book authors and the CaaS provider as outlined in Section \ref{subsec:revenue_sharing}. Only then does the data source become available and ready to be loaded and processed by the CaaS provider. 

Additionally, we generated questions (plain prompts) from areas that vary in proximity to these core topics, allowing for a range of relevance scores in potential retrieved chunks to test how UCOSA would handle them. These areas include explainable AI, security \& privacy, game theory, optimization, and differential privacy.

To gain more confidence with the conducted experiments, we repeated each experiment 100 times, shuffling the prompts each time, and then took the average result.
Our proposed algorithm is designed to be framework-agnostic, with the potential to be implemented within various RAG frameworks and configurations, irrespective of the specific LLM, embedding model, or vector database employed. The framework and configuration described above serve as a representative example to facilitate experiment simulations and demonstrate the algorithm's broader applicability. However, further testing with different LLM implementations would be necessary to fully confirm this generalization.

\subsection{Research Questions}
We perform extensive experiments to answer the following research questions:
\begin{itemize}
    \item How does the performance of UCOSA compare to offline selection (best case) algorithm and ``relevance-greedy'' selection algorithms?
    \item What impact do various allocated budgets have on the performance of UCOSA?    
    \item How do the performance, used budget, and performance-to-budget ratio for LB-CaaS compare to OB-CaaS and RaaS variants?   
\end{itemize}

\subsection{Experimental Scenarios And Analysis}
The key factors that impact the performance $(NEP \times AR)$ of the proposed CaaS variations are the chunk selection strategy, budget type (open or limited), and the allocated budget. For that, we design several experimental scenarios that investigate the impact of each factor on the overall performance. 

\subsubsection{The performance of UCOSA compared to offline and ``relevance-greedy'' selection algorithms}

\begin{figure*}[!t]
    \centering 
    \begin{subfigure}{0.32\textwidth}
        \includegraphics[width=\linewidth]{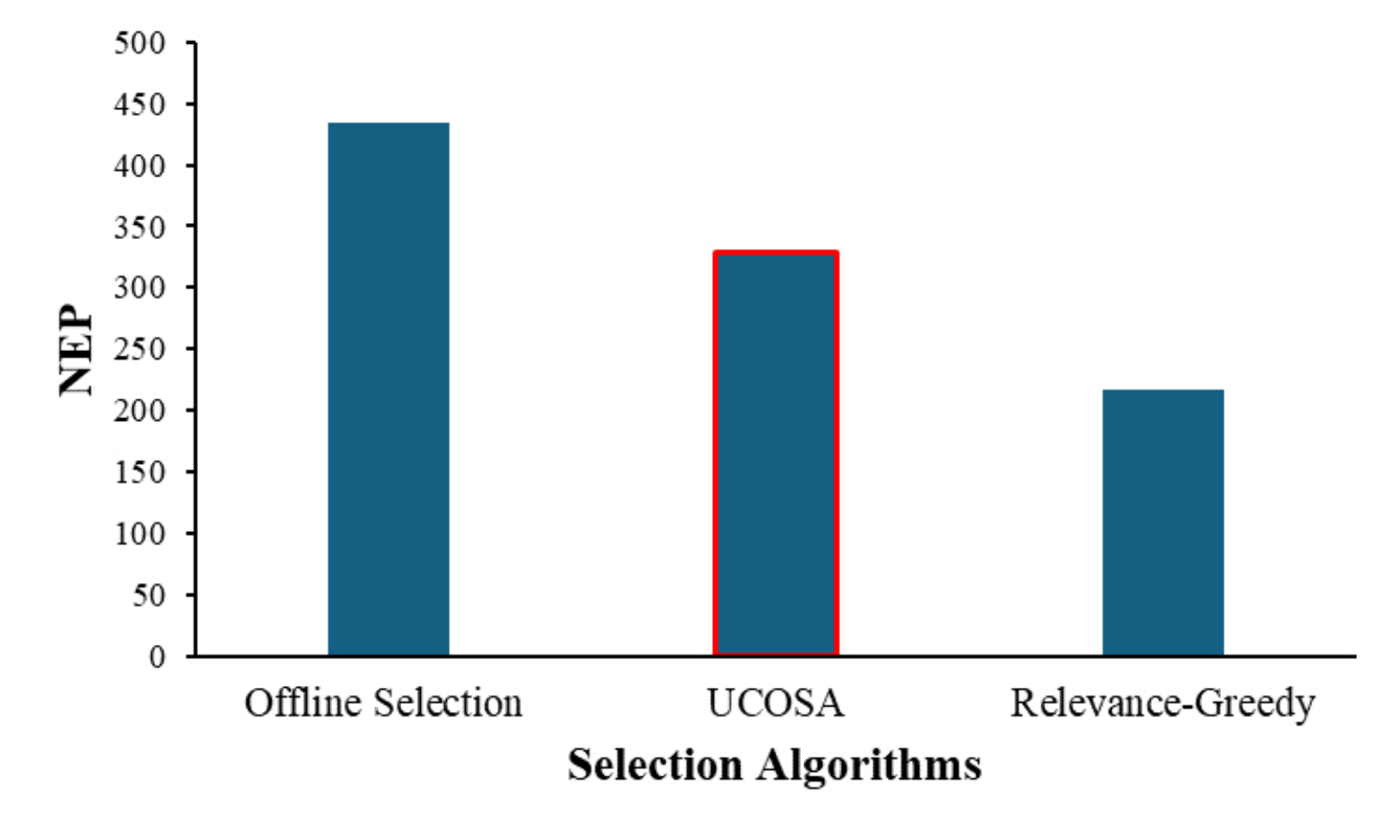}
        \caption{When UCOSA is used, $NEP$ is higher than relevance-greedy selection and competitive with offline selection.}
        \label{fig:number_enriched_prompts(NEP)_UCOSA}
    \end{subfigure}
    \hspace*{\fill}
    \begin{subfigure}{0.32\textwidth}
        \includegraphics[width=\linewidth]{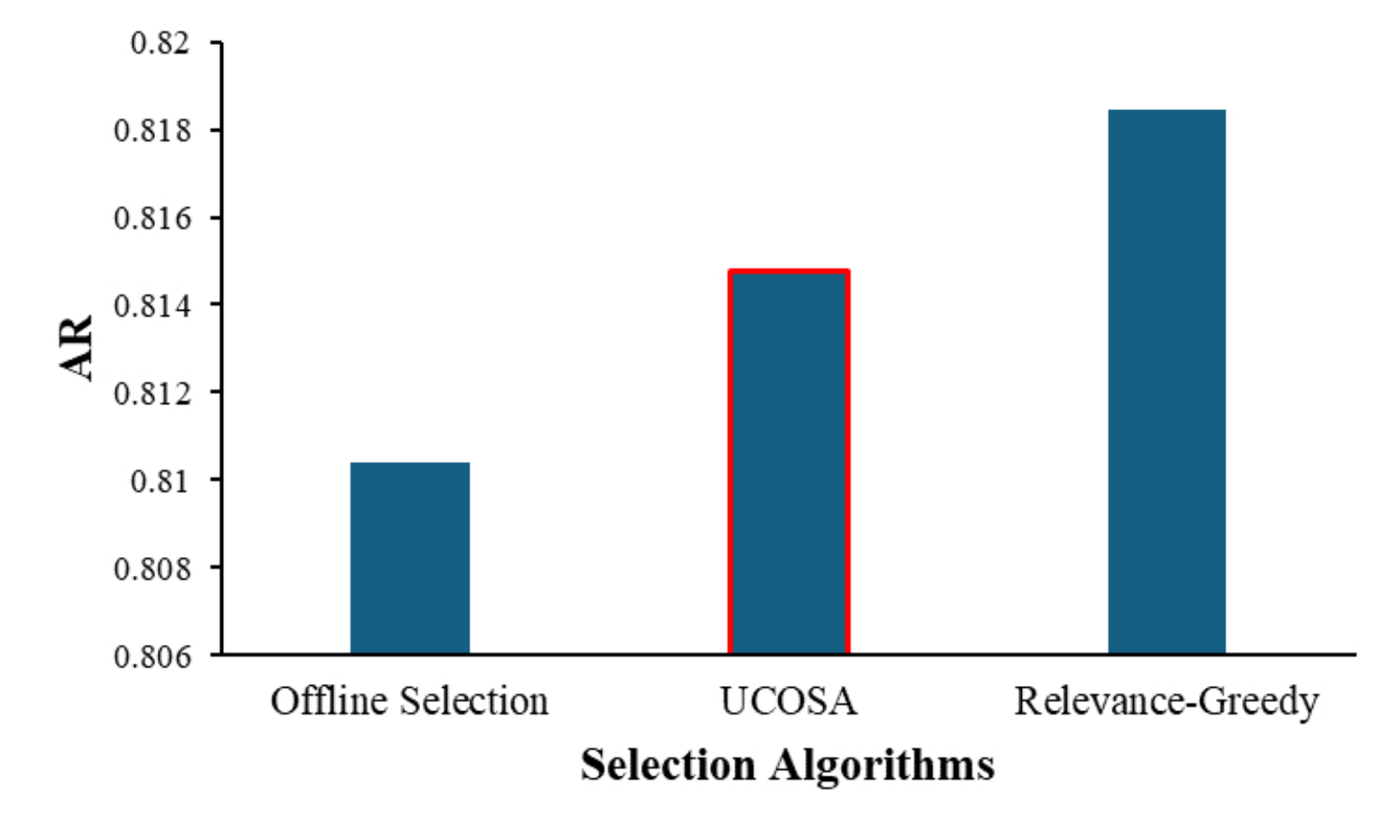}
        \caption{When using UCOSA, $AR$ of the selected chunks is superior to offline selection and competitive to relevance-greedy algorithms.}
        \label{fig:average_relevance(AR)_UCOSA}
    \end{subfigure}
    \hspace*{\fill}
    \begin{subfigure}{0.32\textwidth}
         \includegraphics[width=\linewidth]{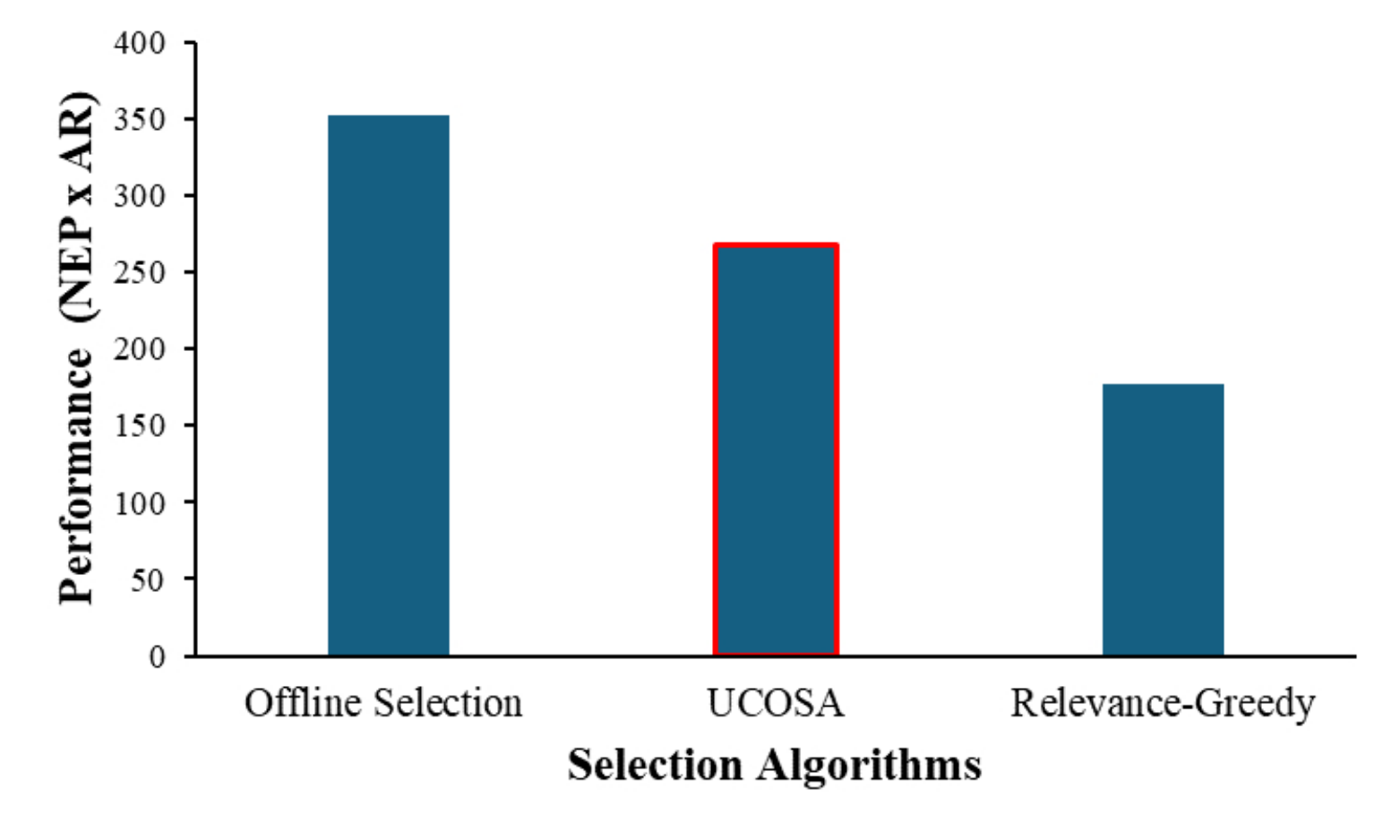}
        \caption{UCOSA outperforms ``relevance-greedy'' selection and is competitive with offline selection algorithms.}
        \label{fig:performance(NEPxAR)_UCOSA}
    \end{subfigure}
    \caption{Performance of UCOSA compared to the offline (best case) and relevance-greedy selection algorithms considering $NEP$, $AR$, and ($NEP x AR$). For the considered performance metric ($NEP x AR$), UCOSA shows superior performance compared to relevance-greedy selection and competitive performance to the offline selection algorithm.}
    \label{fig:performance_UCOSA_compared_offline_and_random_selection_algorithms}
\end{figure*}

This experimental scenario investigates the performance of UCOSA compared to the offline (best case) and ``relevance-greedy'' selection algorithms. Fig. \ref{fig:performance_UCOSA_compared_offline_and_random_selection_algorithms} presents the results of this scenario considering $NEP$, $AR$, and ($NEP \times AR$).

Starting with $NEP$, Fig. \ref{fig:number_enriched_prompts(NEP)_UCOSA} illustrates the number of enriched prompts ($NEP$) when UCOSA is utilized compared to the usage of offline and ``relevance-greedy'' selection algorithms. In terms of $NEP$, which represents the quantity of the target generated output, UCOSA achieves a higher $NEP$ than ``relevance-greedy'' selection and performs competitively with offline selection. 
The competitive performance of UCOSA in terms of $NEP$, compared to the benchmarked offline selection (which is impractical for real-time RAG systems), underscores UCOSA's efficiency and effectiveness in real-time applications. Its ability to achieve competitive $NEP$ in dynamic and unpredictable environments makes UCOSA a strong candidate for RAG systems that require immediate responses.

Moving to $AR$, Fig. \ref{fig:average_relevance(AR)_UCOSA} shows the average relevance scores ($AR$) of the selected chunks when UCOSA is utilized compared to the usage of ``relevance-greedy'' and offline selection algorithms. In terms of $AR$, which represents the quality of the target generated output, UCOSA achieves competitive performance compared to the ``relevance-greedy'' selection algorithm and superior performance compared to offline selection algorithms. 
The ``relevance-greedy'' selection algorithm strives to maximize the $AR$ of the selected chunks regardless of their associated prices. In other words, the ``relevance-greedy'' approach prioritizes relevance scores over cost reduction, potentially sacrificing output quantity. Therefore, it achieves higher $AR$.
Interestingly, the $AR$ for UCOSA is higher than that of offline selection. One interpretation of this observation is that since the NEPs are high when offline selection is utilized, there is a possibility that some of the involved prompts are enriched with chunks of low relevance score values. These low relevance scores, when included in calculating the average, could reduce the $AR$ for offline selection.
UCOSA's ability to maintain high relevance in real-time selection highlights its robustness in dynamic environments.

Concerning UCOSA performance, Fig. \ref{fig:performance(NEPxAR)_UCOSA} illustrates that the performance metric $(NEP \times AR)$ fairly evaluates the performance of UCOSA because it considers both the quality and quantity of the generated output.

For this metric, UCOSA outperforms ``relevance-greedy'' selection by approximately 52\% and achieves around 75\% of the performance of offline selection methods, demonstrating its effectiveness as a competitive, real-time selection method.

Unlike offline selection, which assumes the availability of all prompts beforehand---a scenario not typical in real-world LLM applications---UCOSA operates in real time, dynamically selecting prompts as they arrive. This makes UCOSA particularly valuable for practical applications where prompt streams are unpredictable and not fully available in advance.

\subsubsection{The impact various allocated budgets have on the performance of UCOSA}

In this experimental scenario, we investigate the impact of various allocated budgets on the performance of UCOSA. 

\begin{figure}[]
    \centering 
     \includegraphics[width=\linewidth]{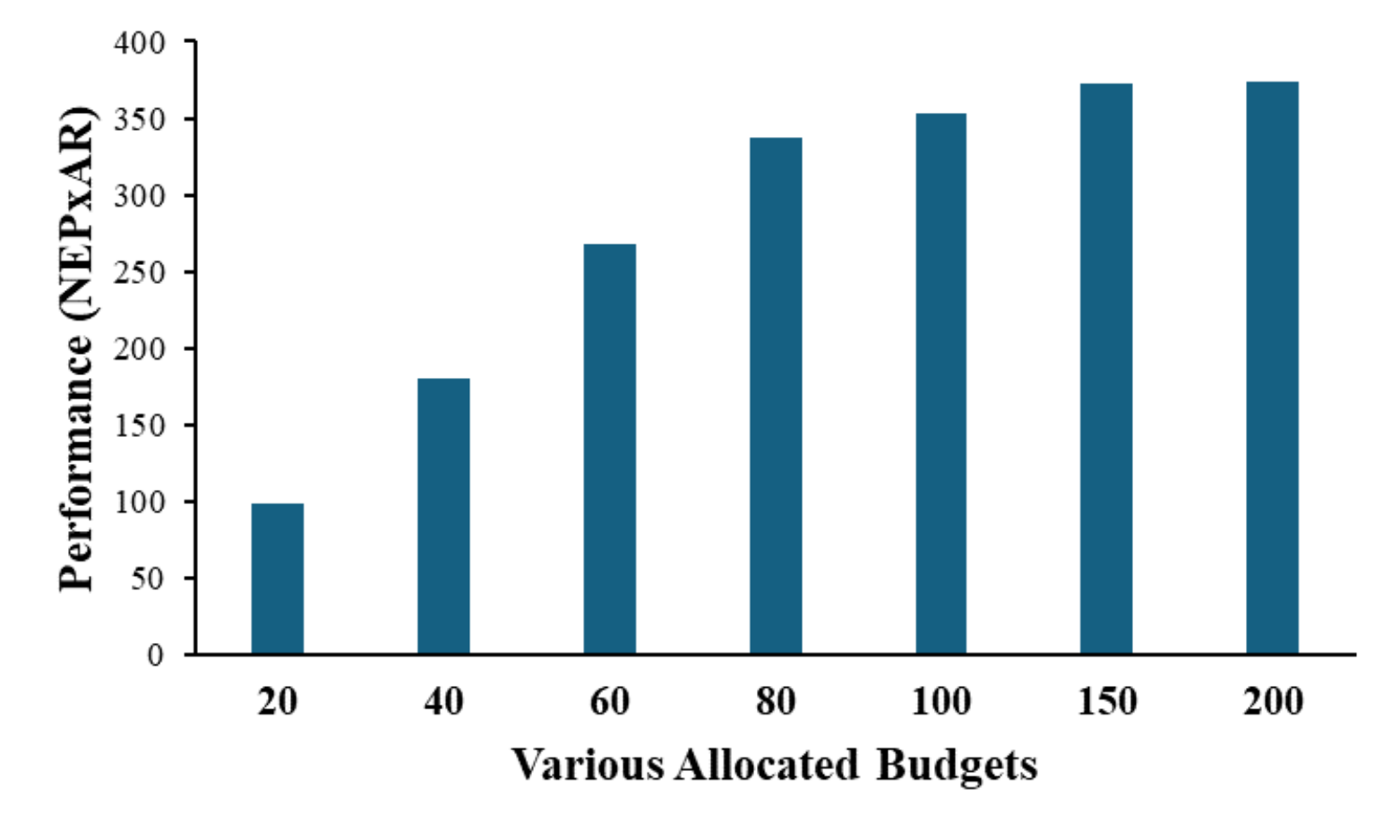}
    \caption{UCOSA's performance improves with increased budget until it plateaus. This plateau indicates the budget required to enrich all plain prompts, marking the conversion from LB-CaaS to OB-CaaS.}
    \label{fig:performance_and_various_budget_allocations}
\end{figure}

Fig. \ref{fig:performance_and_various_budget_allocations} illustrates the impact various allocated budgets have on the performance of UCOSA. UCOSA's performance positively correlates with the increased allocated budget. However, it reaches a point where performance stops increasing no matter what the increased budget is. This point is interpreted as the required budget to enrich all plain prompts. In other words,
the point where LB-CaaS converts to OB-CaaS variant (i.e., $B=\infty$).

\subsubsection{The the performance, used budget, and performance-to-budget ratio for LB-CaaS compared to OB-CaaS and RaaS}

\begin{figure*}[!t]
    \begin{subfigure}{0.32\textwidth}
         \includegraphics[width=\linewidth]{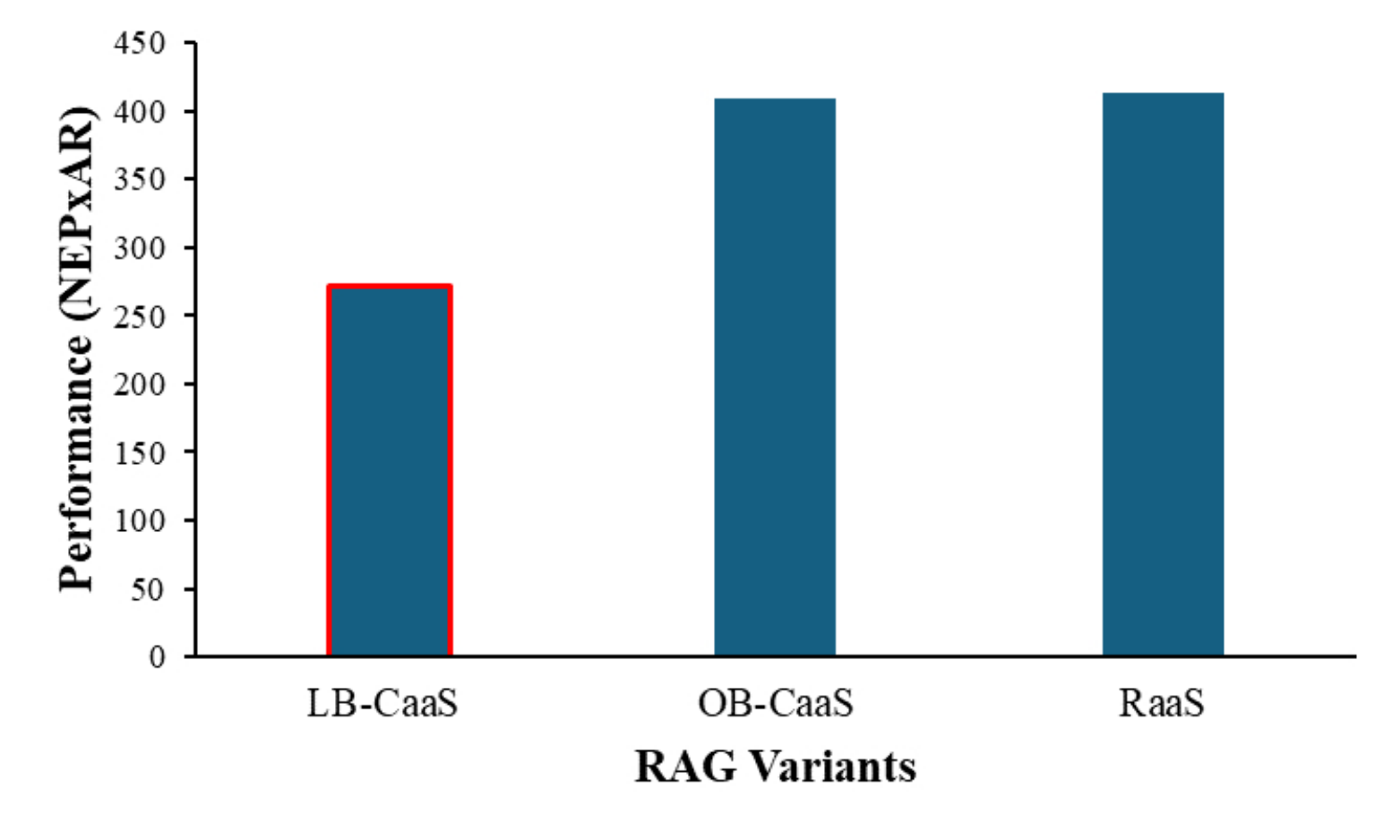}
        \caption{OB-CaaS outperforms LB-CaaS and matches RaaS performance, albeit at a higher cost.}
        \label{fig:peformance_of_OB_CaaS_compared_to_LB_CaaS}
    \end{subfigure}
    \hspace*{\fill}
    \begin{subfigure}{0.32\textwidth}
         \includegraphics[width=\linewidth]{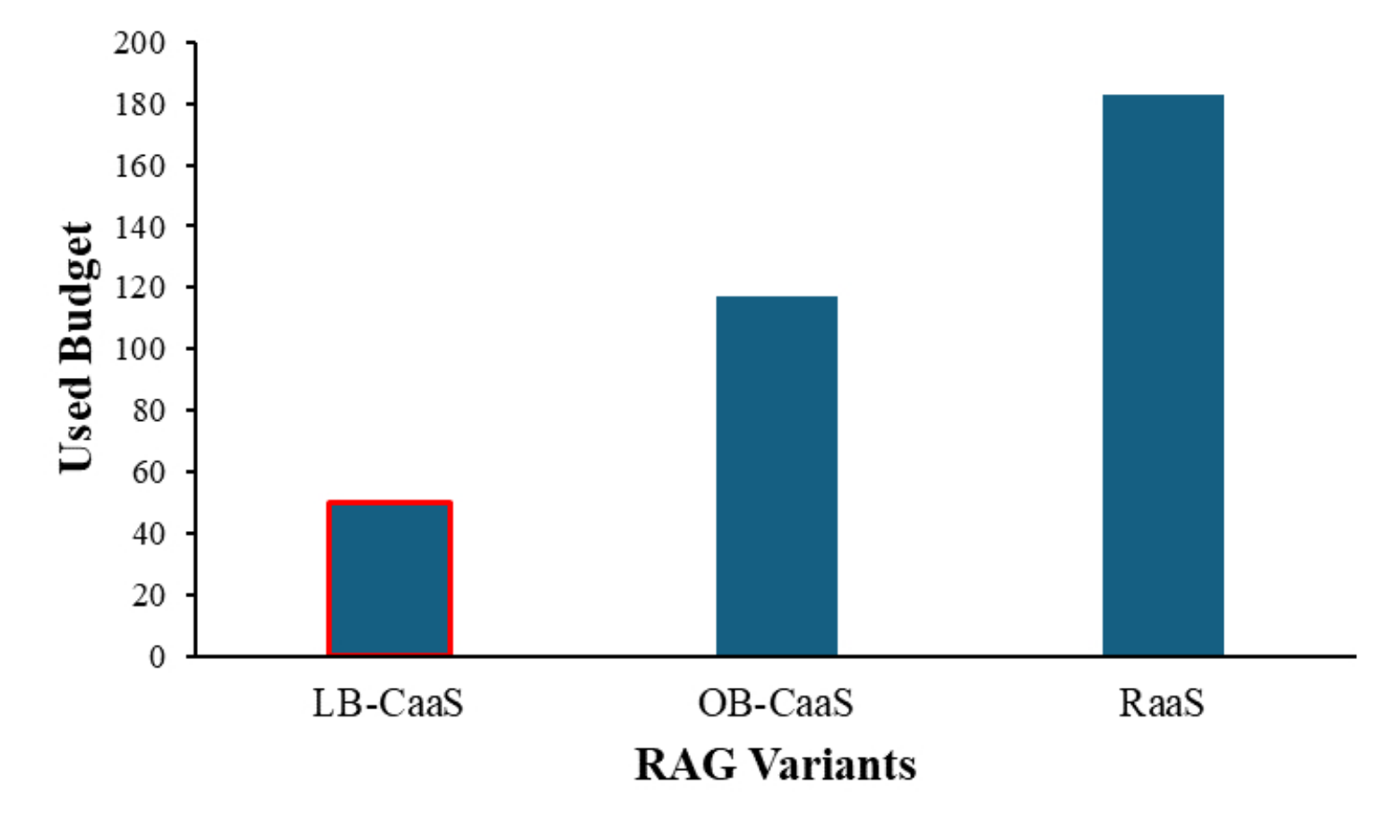}
        \caption{The budget of LB-CaaS compared to OB-CaaS and RaaS (lower is better). LB-CaaS's budget aligns with its performance shown in Fig.~\ref{fig:peformance_of_OB_CaaS_compared_to_LB_CaaS}}
        \label{fig:Used_Budget_of_OB_CaaS_compared_to_LB_CaaS}
    \end{subfigure}
    \hspace*{\fill}
    \begin{subfigure}{0.32\textwidth}
         \includegraphics[width=\linewidth]{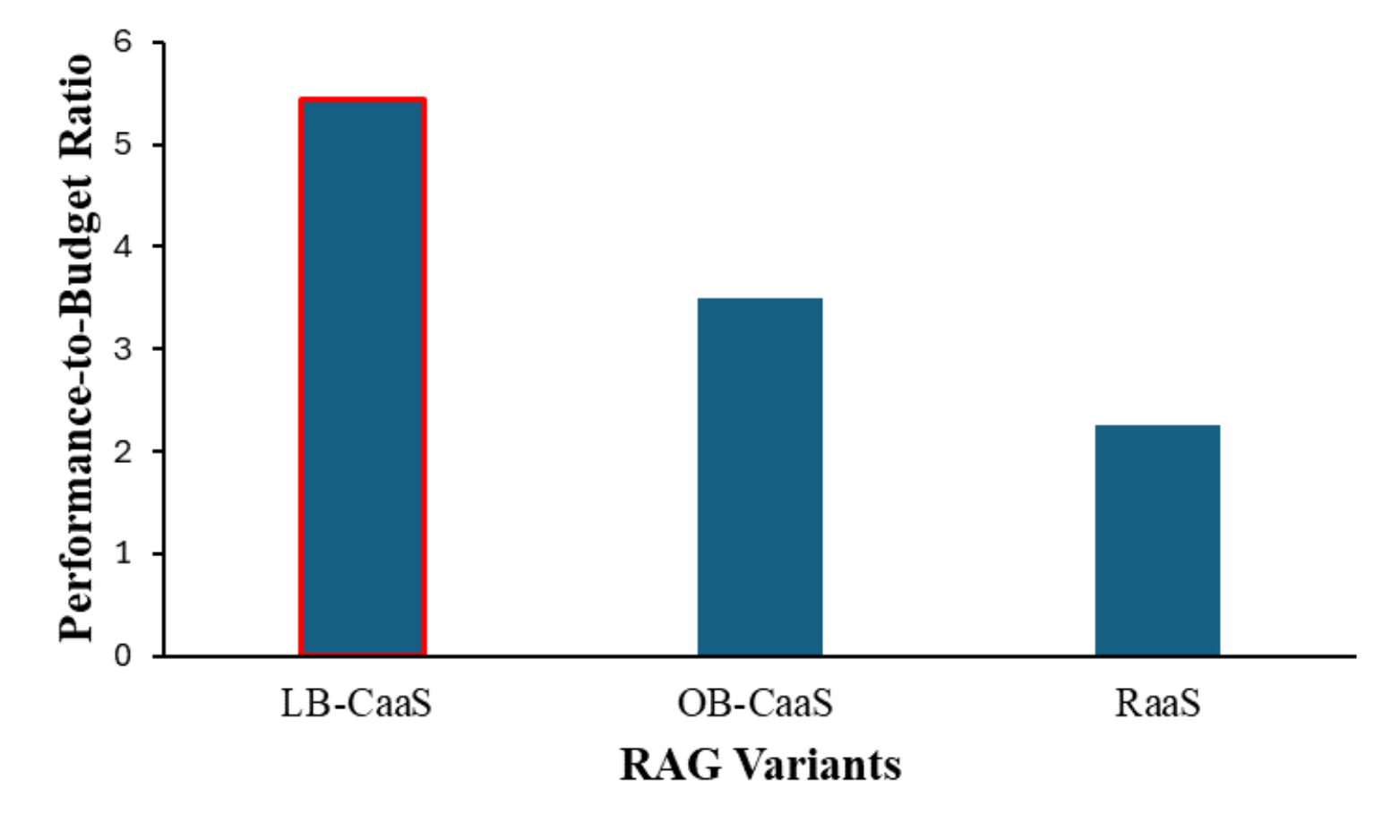}
        \caption{Performance-to-budget ratio shows higher budget utilization for LB-CaaS (enabled by UCOSA) compared to OB-CaaS and RaaS.}
        \label{fig:Ratios_of_UsedBudget_and_Performace_for_OB-CaaS_to_LB-CaaS}
    \end{subfigure}
\caption{The performance, used budget, and the performance-to-budget ratio of LB-CaaS compared to the OB-CaaS and RaaS variants.
LB-CaaS exhibits the highest performance-to-budget ratio, indicating its superior efficiency in budget utilization compared to OB-CaaS and RaaS. OB-CaaS demonstrates intermediate efficiency, while RaaS shows the lowest efficiency in budget utilization.
}
\label{fig: LB_CaaS_vs_OB-CaaS}
\end{figure*}

Fig. \ref{fig: LB_CaaS_vs_OB-CaaS} illustrates an experimental scenario that compares the performance, used budget, and performance-to-budget ratio of LB-CaaS with the OB-CaaS and RaaS variants.

Unlike OB-CaaS and LB-CaaS variants, where the cost is chunk-dependent, the cost of RaaS is calculated for the submitted prompts. For the sake of a fair comparison, we consider the cost of each submitted prompt to RaaS as the average cost of the chunks in OB-CaaS and LB-CaaS variants.
Firstly, Fig. \ref{fig:peformance_of_OB_CaaS_compared_to_LB_CaaS} demonstrates the performance of LB-CaaS compared to OB-CaaS and RaaS (the higher, the better). 
LB-CaaS shows lower performance than OB-CaaS and RaaS. However,
OB-CaaS and RaaS achieve this superiority at the expense of a higher cost, i.e., a significantly higher budget as demonstrated in Fig. \ref{fig:Used_Budget_of_OB_CaaS_compared_to_LB_CaaS}.

Secondly, Fig. \ref{fig:Used_Budget_of_OB_CaaS_compared_to_LB_CaaS} shows the used budget of LB-CaaS compared to OB-CaaS and RaaS (the lower, the better). The used budget of LB-CaaS is proportional to the associated performance illustrated in Fig. \ref{fig:peformance_of_OB_CaaS_compared_to_LB_CaaS} and LB-CaaS is the most cost-efficient variant as demonstrated in Fig. \ref{fig:Ratios_of_UsedBudget_and_Performace_for_OB-CaaS_to_LB-CaaS}. Furthermore, we observe that although the used budget for OB-CaaS is less than that of RaaS, the associated performance of both is similar, highlighting the cost efficiency of OB-CaaS.

Lastly, Fig. \ref{fig:Ratios_of_UsedBudget_and_Performace_for_OB-CaaS_to_LB-CaaS} 
evaluates the performance-to-budget ratio for each RAG variant.
LB-CaaS and OB-CaaS achieve higher performance-to-budget ratios, 140\% and 86\%, respectively, compared to RaaS, indicating their superior efficiency in budget utilization.
In particular, LB-CaaS exhibits the highest performance-to-budget ratio, indicating its superior efficiency in budget utilization compared to OB-CaaS and RaaS. OB-CaaS demonstrates intermediate efficiency, while RaaS shows the lowest efficiency in budget utilization. 
This analysis underscores the practicality of LB-CaaS. By balancing the cost with performance, LB-CaaS emerges as a viable and cost-effective RAG variant.

\subsubsection{Human-in-th-Loop Evaluation}
To complement the simulation-based results with human-in-the-loop verification, we assessed how chunk-level pricing and prompt enrichment influence the actual user experience by incorporating human experts into the evaluation process. In particular, three human experts (information retrieval researchers) investigated the chunks associated with 50 randomly sampled prompts, comparing those selected by our proposed UCOSA (the enabler of LB-CaaS) against chunks selected by the baseline approaches (Offline and Relevance-Greedy). The findings of the experts emphasized the simulated results: chunks selected by our proposed work were found to be sufficiently informative and relevant to the served prompts while also being more cost-effective (associated with less price). On the other hand, chunks selected by the baseline approaches were found to be of comparable informativeness but with higher prices. These results emphasize that UCOSA effectively balances utility (relevance) and cost (price) in a way that positively influences the actual user experience and information satisfaction.

\subsection{Insights and Lessons learned}
\label{lab:lessens_learned}

\begin{itemize}
\item
Relying solely on the relevance or price might lead to ineffective generated output. For example, relying on chunk relevance might lead to serving fewer prompts when the top relevant ones are associated with higher prices. There could be cheaper chunks with a degree of relevance scores that are sufficient to generate quality output. However, these chunks are neglected when utilizing relevance as the only selection metric. Likewise, if we only rely on the price for the selection of the chunks, the process tends to select the cheaper chunks even if they are associated with lower relevance scores that are insufficient to generate quality output. Therefore, striking a balance between relevance and price is a prudent strategy.
\item OB-CaaS can be extended to various enhanced variants besides the proposed LB-CaaS. For example, a variant optimized for the cost (OFC-CaaS) and another optimized for the relevance (OFR-CaaS). OFC-CaaS would be suitable for the use cases that prioritize the quantity of enriched prompts over their quality. In contrast, the OFR-CaaS variant is suitable for use cases that prioritize the quality of enriched prompts over their quantity. Our proposed variant LB-CaaS, on the other hand, strikes a balance between both quantity and quality, making it ideal for most real-world scenarios where both aspects are essential.
 \item An edge case of the allocated budgets in the LB-CaaS variant could be a budget that is sufficient for handling all prompts in the stream ($B=\infty$), which converts it to OB-CaaS. 
\item The proposed OB-CaaS variant involves post-calculation of the total price associated with the utilized chunks, implying less control over the budget. In contrast, the LB-CaaS variant operates with a pre-defined budget, implying more control over the budget.
\end{itemize}

\section{Conclusion and Future Work} \label{sec:Conclusion_and_future_work}
We propose two RAG variants to address the standard RAG-as-a-Service (RaaS) limitations. Open-Budget CaaS (OB-CaaS) is the basic variant that renders standard RaaS cost-effective, allowing accessibility to broader users interested in utilizing RAG. The second variant is Limited-Budget CaaS (LB-CaaS). It is a cost-aware variant extending the accessibility of the OB-CaaS to serve wider RAG-interested users. We investigated the performance of our proposed RAG variants by conducting extensive experiments, demonstrating their efficacy and proving the optimality of UCOSA (the enabler of LB-CaaS) mathematically and empirically.
In future work, we plan to investigate the feasibility of integrating the LB-CaaS model with a hallucination detection technique, enabling the LB-CaaS to function as a hallucination-detection-driven model to further enhance its cost-effectiveness. In particular, we aim to explore the cost reduction compared to the expected hallucination detection overhead.

Inspired by the typical FL setting, as future work, we suggest a variant of FL that leverages UCOSA to mitigate privacy concerns, while enabling CaaS-like functionality. This variant aims to avoid sharing the whole data sources and only sharing minimal data (one record for each process) when necessary. It would follow a two-stage selection approach. In the first stage, a copy of the served prompt is shared with all decentralized locations, where UCOSA is deployed and utilized to select the most feasible chunk (considering relevance and price) as a candidate and send it to the centralized location for further processing. In the second stage, the centralized location utilizes UCOSA to process the candidate chunks to make a final selection decision by selecting the most feasible ones conditioned by the remaining budget.
Another future direction is to develop an online algorithm that targets dynamic pricing scenarios, where the cost of a chunk may change over time. In addition, we plan to conduct larger-scale human-in-loop evaluation in future research, including user-centered experiments with practitioners and general users to quantify satisfaction, usability, and decision-making impact.


\appendices
\section{UCOSA's Competitive Ratio Analysis}\label{sec:Proof}
In this section, we show that UCOSA achieves a competitive ratio of $\ln (U/L) + 2$. Moreover, we show that the best competitive ratio achieved by any online algorithm cannot be lower than $\ln (U/L) + 1$. The competitive ratio is a measure of the worst-case performance of the online algorithm compared to the optimal solution. In our work, let $\mathcal{R}_{UCOSA}$ be the total Relevance score achieved by UCOSA, and let $\mathcal{R}_{OPT}$ be the total Relevance score achieved by the optimal solution (the optimization formulation in the main manuscript, Section IV-B), then we show that 

\begin{displaymath}
	\frac{\mathcal{R}_{OPT}}{\ln (U/L) + 2} \leq \mathcal{R}_{UCOSA} \leq \mathcal{R}_{OPT}
\end{displaymath}

The proof starts by examining the sets of prompts enriched with a selected chunk $j$ by the offline algorithm and UCOSA, denoted by $S_j^*$ and $S_j$, respectively. The set $S_j^*$ is partitioned into subsets. Leveraging the properties of $\Psi(z_i)$ and the selection condition of UCOSA (Line 7 of Algorithm 1), we bound the relevance score achieved by each subset compared to the relevance score achieved by UCOSA. Taking the ratio of $\mathcal{R}_{OPT}/\mathcal{R}_{UCOSA}$ over all chunks yield the competitive ratio result.

Before delving into the mathematical proof, we note the following properties of the function $\Psi(z_i)$ used in UCOSA:
\begin{itemize}
	\item When $z_i \in [0, \frac{1}{1 + \ln (U/L)}]$, the value of $\Psi(z_i) \leq L$. Therefore, all prompts will be enriched with a chunk.
	\item When $z_i = 1$, the value of $\Psi (z_i) = U$. From the algorithm, no additional prompts will be enriched, and the budget is not violated.
	\item $\Psi(z_i)$ is a monotonically increasing function of $z_i$.
\end{itemize}

We now prove the competitive ratio of UCOSA.
\subsubsection{Proof of Competitive Ratio}
The proof is as follows.
\begin{proof}
	Fix an input sequence $\zeta$. Let $OPT(\zeta)$ and $A(\zeta)$ be the total Relevance score obtained by the optimal offline and UCOSA, respectively. Suppose that when the UCOSA terminates, the fraction of budget consumed to use chunk $j$ is $Z_j$, and let $B_j = Z_j B$. Let $S_j^* , S_j$ be the set of prompts enriched with chunk $j$ by the optimal offline and the UCOSA, respectively.
	
	We partition the set $S_j^*$ into 3 subsets:
	\begin{itemize}
		\item The prompts enriched by both offline and UCOSA with chunk $j$. Denote this by $D_j^* = S_j^* \cap S_j$
		\item The prompts enriched by the offline with chunk $j$, but not enriched by UCOSA. Denote this by $X_j^* = S_j^* \backslash S_j$.
		\item The prompts enriched by the offline with chunk $j$, but enriched with different chunk by UCOSA. Denote this by $Y_j^*$.
	\end{itemize}
	
	For the prompts in $D_j^*$, since it was enriched by UCOSA, then $R_{ij} \geq P_{ij}\Psi(z_i), \forall i \in D_j^*$. The total Relevance score $v(D_j^*)$ of these prompts is:
	\begin{eqnarray}
		v(D_j^*) &= \sum_{i \in D_j^*} R_{ij} \nonumber\\
		&\geq \sum_{i \in D_j^*} P_{ij}\Psi(z_i) \label{eq:eq1}
	\end{eqnarray}
	Since $\Psi(z_i)$ is monotonically increasing, The right hand side of Equation~\eqref{eq:eq1}: 
	\begin{equation}
		\sum_{i \in D_j^*} P_{ij}\Psi(z_i) \leq \sum_{i \in D_j^*} P_{ij}\Psi(Z_j) \leq B\Psi(Z_j) \label{eq:D}
	\end{equation}
	
	For the prompts in $X_j^*$, since these were not enriched by UCOSA, then $R_{ij} \leq P_{ij}\Psi(z_i), \forall i \in X_j^*$. Therefore:
	\begin{eqnarray}
		\sum_{i \in X_j^*} R_{ij} &\leq \sum_{i \in X_j^*} P_{ij}\Psi(z_i) \nonumber\\
		&\leq \sum_{i \in X_j^*} P_{ij}\Psi(Z_j) \nonumber\\
		&\leq \Psi(Z_j)(B - B_j) \label{eq:X}
	\end{eqnarray}
	
	For the prompts in $Y_j^*$, UCOSA enriched the prompt with a different chunk to achieve a higher score. Therefore:
	\begin{displaymath}
		v_{off}(Y_j^*) \leq v_{UCOSA}(Y_j^*) \leq v(S_j), \forall j
	\end{displaymath} 
	where $v_{off}(Y_j^*)$ and $v_{UCOSA}(Y_j^*)$ are the sum of Relevance scores of the set $Y_j^*$ achieved by the optimal offline and UCOSA, respectively. Therefore:
	\begin{equation}
		\sum_j v_{off}(Y_j^*) \leq \sum_j v(S_j) = A(\zeta) \label{eq:Y}
	\end{equation}
	
	There are also the prompts enriched by UCOSA with chunk $j$, but not enriched by the offline algorithm (\emph{i.e.,} the set $S_j \backslash S_j^*$). For these prompts, we have:
	\begin{equation}
		v(S_j \backslash S_j^*) = \sum_{i \in S_j \backslash S_j^*} R_{ij} \geq \sum_{i \in S_j \backslash S_j^*}P_{ij}\Psi(z_i) \label{eq:G}
	\end{equation}
	
	Note that $OPT(\zeta) = \sum_j \big(v(D_j^*) + v(X_j^*) + v_{off}(Y_j^*)\big)$, and $A(\zeta) = \sum_j \big(v(D_j^*) + v(S_j \backslash S_j^*)\big)$. The competitive ratio is:
	\begin{eqnarray}
		\frac{OPT(\zeta)}{A(\zeta)} &= \frac{\sum_j \big(v(D_j^*) + v(X_j^*) + v_{off}(Y_j^*)\big)}{\sum_j \big(v(D_j^*) + v(S_j \backslash S_j^*)\big)} \nonumber \\
		&\leq \frac{\sum_j \big(v(D_j^*) + v(X_j^*)\big)}{A(\zeta)} + \frac{A(\zeta)}{A(\zeta)} \nonumber \\
		&\leq \frac{\sum_j \big[\sum_{i \in D_j^*} \big(P_{ij}\Psi(Z_j)\big) + \Psi(Z_j)(B - B_j) \big]}{\sum_j \big[\sum_{i \in D_j^*} \big(P_{ij}\Psi(Z_j)\big) + v(S_j \backslash S_j^*) \big]} + 1 \nonumber\\
		&\leq \frac{\sum_j \big[\Psi(Z_j)B_j + \Psi(Z_j)(B - B_j) \big]}{\sum_j \big[\sum_{i \in D_j^*} \big(P_{ij}\Psi(Z_j)\big) + v(S_j \backslash S_j^*) \big]} + 1 \nonumber \\
		&\leq \frac{\sum_j \Psi(Z_j)B}{\sum_j \big( \sum_{i \in S_j} \Psi(z_i)P_{ij} \big)} + 1\nonumber \\
		&\leq \frac{\sum_j \Psi(Z_j)}{\sum_j \big( \sum_{i \in S_j} \Psi(z_i)\Delta z_i \big)} + 1 \label{eq:1}
	\end{eqnarray}
	Where the first inequality is obtained by substituting Equation~\eqref{eq:Y}. The second inequality is obtained by substituting Equation~\eqref{eq:X}, the third inequality is obtained by substituting Equation~\eqref{eq:D}, and the last inequality is obtained by substituting $\Delta z_i = P_{ij}/B$.
	
	For the denominator of Equation~\eqref{eq:1}, we approximate the summation via an integration (since we already assumed that $P_{ij}/B \leq \epsilon$, where $\epsilon \rightarrow 0$). Therefore, for every chunk $j$ we have:
	\begin{eqnarray}
		\sum_{i \in S_j} \Psi(z_i)\Delta z_i &\approx \int_{0}^{Z_j} \Psi(z_i)dz_i \nonumber\\
		&= \int_{0}^{g} Ldz_i + \int_{g}^{Z_j} \Psi(z_i)dz_i \nonumber \\
		&= gL + \frac{L}{e}\frac{(Ue/L)^{Z_j} - (Ue/L)^{g}}{\ln (Ue/L)} \nonumber\\
		&= \frac{L}{e} \frac{(Ue/L)^{Z_j}}{\ln (Ue/L)} \nonumber \\
		&= \frac{\Psi(Z_j)}{\ln (U/L) + 1} \label{eq:2}	  
	\end{eqnarray}
	Where $g = \frac{1}{1 + \ln (U/L)}$.
	
	Substituting Equation~\eqref{eq:2} in Equation~\eqref{eq:1}, we have:
	$\frac{OPT(\zeta)}{A(\zeta)} \leq \frac{\sum_j \Psi(Z_j)}{\frac{\Psi(Z_j)}{\ln (U/L) + 1}} + 1 \leq \ln (U/L) + 2$.

	Since each of the ratios $\frac{\Psi(Z_j)}{\sum_{i \in S_j} \Psi(z_i)\Delta z_i}$ is less than $\frac{1}{\ln (U/L) +1}$, the last inequality follows.
	
\end{proof}

\subsubsection{Lower Bound Proof}
To prove the lower bound on the competitive ratio, we provide an instance of the problem on which no online algorithm can achieve a better competitive ratio than $\ln(U/L) + 1$.

UCOSA presented above is a deterministic algorithm (\emph{i.e.,} given the same input sequence, UCOSA will produce the exact same output). To prove the competitive ratio, we use Yao's principle. Yao's principle states that deterministic algorithms performs worse than randomized algorithms against an oblivious adversary (one who does not know the outcomes of the random process in the randomized algorithm). Therefore, a lower bound on the randomized algorithm is also a lower bound on the deterministic algorithm.

For any input distribution $\mathcal{F}$ and $\gamma$-competitive randomized algorithm $\mathcal{A}$, we will show that:
\begin{displaymath}
	\frac{1}{\gamma} \leq \min_{\sigma} \frac{\mathbb{E}[\mathcal{A}(\sigma)]}{OPT(\sigma)} \leq \max_{deterministic A} \mathbb{E}_{\sigma \leftarrow \mathcal{F}} \big[ \frac{A(\sigma)}{OPT(\sigma)} \big]
\end{displaymath}

To prove the lower bound, we specify a distribution $\mathcal{F}$ such that:
\begin{equation}
	\max_{deterministic A} \mathbb{E}_{\sigma \leftarrow \mathcal{F}} \big[ \frac{A(\sigma)}{OPT(\sigma)}\big] \leq \frac{1}{\ln(U/L) +1}
\end{equation}

The instance of the problem is defined as follows:
\begin{itemize}
	\item The instance has one chunk.
	\item Fix a parameter $\eta > 0$. Let $k$ be the largest integer such that $(1 + \eta)^k \leq U/L$. Therefore, $k = \lfloor \frac{\ln(U/L)}{\ln(1+\eta)} \rfloor$.
	\item The support of the input distribution $\mathcal{F}$ consist of the instances $I_0, I_1, \ldots, I_k$, where $I_0$ consists of $B$ identical prompts each with price 1 and Relevance score of $L$, and $I_{m+1}$ consists of $I_m$ followed by additional $B$ prompts each with price 1 and a Relevance score $(1+\eta)^{m+1}L$. This means that $I_m \subset I_{m+1}$.
	\item The probabilities of the instances $I_0, I_1, \ldots, I_k$ are defined as follows: $p_k = \frac{1+\eta}{(k+1)\eta + 1}$, and $p_m = \frac{\eta}{(k+1)\eta + 1}, \forall m \in \{0, 1, \ldots, k-1\}$. Note that $\sum_{m=0}^{k} p_m = 1$. 
	\item Define $H = (k+1)\eta + 1$.  
\end{itemize}

\begin{proof}
	Let $b_m$ be the fraction of the budget used for prompts with profits $(1+\eta)^m L$, $m = 0, 1, \ldots, k$. We note that $\sum_{m=0}^{k} b_m \leq 1$ since a proper algorithm will not use more than the available budget. Now, we have:
	
	\begin{eqnarray}
		\mathbb{E}_{\sigma \leftarrow \mathcal{F}} \big[ \frac{A(\sigma)}{OPT(\sigma)} \big]&= \sum_{i=0}^{k} \frac{p_i \sum_{m=0}^{i} (1+\eta)^j b_m}{(1+\eta)^i} \nonumber \\
		&= \sum_{m=0}^{k}b_m \sum_{i=m}^{k} p_i (1+\eta)^{m-i} \nonumber 
	\end{eqnarray}
	
	For any $m$:
	\begin{eqnarray}
		\sum_{i=m}^{k} p_i (1+\eta)^{m-i} &= p_k (1+\eta)^{m-k} + \sum_{i=m}^{k-1} p_i (1+\eta)^{m-1} \nonumber \\
		&= \frac{(1+\eta)^{m-k+1}}{H} + \frac{\eta}{H} \sum_{i=m}^{k-1}(1+\eta)^{m-i} \nonumber \\
		&= \frac{(1+\eta)^{m-k+1}}{H} + \frac{\eta}{H} \frac{(1+\eta) - (1+\eta)^{m-k+1}}{\eta} \nonumber \\
		&= \frac{1+\eta}{H} = \frac{1+\eta}{(k+1)\eta + 1} \nonumber
	\end{eqnarray}
	
	where the second equality comes from substituting the values of $p_m$, and the last equality comes from substituting the value of $H$.
	
	Therefore, we get:
	\begin{eqnarray}
		\mathbb{E}_{\sigma \leftarrow \mathcal{F}} \big[ \frac{A(\sigma)}{OPT(\sigma)} \big] &= \frac{1+\eta}{(k+1)\eta + 1}\sum_{m=0}^{k}b_m \nonumber \\
		&\leq \frac{1+\eta}{(k+1)\eta + 1} \nonumber \\
		&\leq \frac{1+\eta}{\eta \frac{\ln(U/L)}{\ln(1+\eta)} + 1} \nonumber 
	\end{eqnarray}
	where the first inequality comes from $\sum_{m=0}^{k}b_m \leq 1$, and the second inequality comes from substituting the value of $k = \lfloor \frac{\ln(U/L)}{\ln(1+\eta)} \rfloor$.
	
	By taking the limit as $\eta \rightarrow 0$, and noting that $\lim_{\eta \rightarrow 0} \frac{\eta}{\ln(1+\eta)} =1$, we complete the proof.
\end{proof}
\label{sec:appendix}

\section*{Acknowledgment}
Research reported in this publication was supported by the Qatar Research Development and Innovation Council grant \# ARG01-0525-230348. The content is solely the responsibility of the authors and does not necessarily represent the official views of Qatar Research Development and Innovation Council.

\bibliographystyle{IEEEtran}
\bibliography{IEEEabrv,Bib}

@www{llamaindex_2024,
author = {Llamaindex},
title = {{Lamaindex RAG Framework}},
url = {https://docs.llamaindex.ai/en/stable/},
note = {Accessed: 2024-05-06}
}

@www{openai_chatgpt3.5,
author = {OpenAI},
title = {{ChatGPT 3.5}},
url = {https://www.chatgpt.com/},
note = {Accessed: 2024-05-06}
}

@book{warr2019strengthening,
  title={Strengthening deep neural networks: Making AI less susceptible to adversarial trickery},
  author={Warr, Katy},
  year={2019},
  publisher={O'Reilly Media}
}

@book{chen2022adversarial,
  title={Adversarial robustness for machine learning},
  author={Chen, Pin-Yu and Hsieh, Cho-Jui},
  year={2022},
  publisher={Academic Press}
}

@article{brown2020language,
  title={Language models are few-shot learners},
  author={Brown, Tom and Mann, Benjamin and Ryder, Nick and Subbiah, Melanie and Kaplan, Jared D and Dhariwal, Prafulla and Neelakantan, Arvind and Shyam, Pranav and Sastry, Girish and Askell, Amanda and others},
  journal={Advances in neural information processing systems},
  volume={33},
  pages={1877--1901},
  year={2020}
}

@article{achiam2023gpt,
  title={Gpt-4 technical report},
  author={Achiam, Josh and Adler, Steven and Agarwal, Sandhini and Ahmad, Lama and Akkaya, Ilge and Aleman, Florencia Leoni and Almeida, Diogo and Altenschmidt, Janko and Altman, Sam and Anadkat, Shyamal and others},
  journal={arXiv preprint arXiv:2303.08774},
  year={2023}
}

@article{reid2024gemini,
  title={Gemini 1.5: Unlocking multimodal understanding across millions of tokens of context},
  author={Reid, Machel and Savinov, Nikolay and Teplyashin, Denis and Lepikhin, Dmitry and Lillicrap, Timothy and Alayrac, Jean-baptiste and Soricut, Radu and Lazaridou, Angeliki and Firat, Orhan and Schrittwieser, Julian and others},
  journal={arXiv preprint arXiv:2403.05530},
  year={2024}
}

@inproceedings{guu2020retrieval,
  title={Retrieval augmented language model pre-training},
  author={Guu, Kelvin and Lee, Kenton and Tung, Zora and Pasupat, Panupong and Chang, Mingwei},
  booktitle={International conference on machine learning},
  pages={3929--3938},
  year={2020},
  organization={PMLR}
}

@article{lewis2020retrieval,
  title={Retrieval-augmented generation for knowledge-intensive nlp tasks},
  author={Lewis, Patrick and Perez, Ethan and Piktus, Aleksandra and Petroni, Fabio and Karpukhin, Vladimir and Goyal, Naman and K{\"u}ttler, Heinrich and Lewis, Mike and Yih, Wen-tau and Rockt{\"a}schel, Tim and others},
  journal={Advances in Neural Information Processing Systems},
  volume={33},
  pages={9459--9474},
  year={2020}
}

@misc{nuclia_RaaS,
  author = {{Nuclia}},
  title = {{RAG-as-a-Service}},
  url = {https://nuclia.com/rag-as-a-service/},
  note = {[Online; accessed 01-June-2024]},
  year = {2024}
}

@misc{geniusee_RaaS,
  author = {{Geniusee}},
  title = {{RAG-as-a-Service}},
  url = {https://geniusee.com/retrieval-augmented-generation},
  note = {[Online; accessed 02-June-2024]},
  year = {2024}
}

@misc{amazon_bedrock_RaaS,
  author = {{Amazon Bedrock}},
  title = {{RAG-as-a-Service}},
  url = {https://aws.amazon.com/bedrock/},
  note = {[Online; accessed 02-June-2024]},
  year = {2024}
}

@article{gao2023retrieval,
  title={Retrieval-augmented generation for large language models: A survey},
  author={Gao, Yunfan and Xiong, Yun and Gao, Xinyu and Jia, Kangxiang and Pan, Jinliu and Bi, Yuxi and Dai, Yi and Sun, Jiawei and Wang, Haofen},
  journal={arXiv preprint arXiv:2312.10997},
  year={2023}
}

@inproceedings{chen2024benchmarking,
  title={Benchmarking large language models in retrieval-augmented generation},
  author={Chen, Jiawei and Lin, Hongyu and Han, Xianpei and Sun, Le},
  booktitle={Proceedings of the AAAI Conference on Artificial Intelligence},
  volume={38},
  number={16},
  pages={17754--17762},
  year={2024}
}

@www{vectara_RAGaaS,
author = {Vectara_RAG-as-a-Service},
title = {{Vectara RAG-as-a-Service }},
url = {https://vectara.com/retrieval-augmented-generation/},
note = {Accessed: 2024-08-07}
}

@www{LlamaIndex_Cloud,
author = {LlamaIndex Cloud},
title = {{LlamaIndex Cloud }},
url = {https://docs.llamaindex.ai/en/stable/llama_cloud/},
note = {Accessed: 2024-08-07}
}

@www{LlamaIndex_Embedding,
author = {LlamaIndex},
title = {{LlamaIndex Embedding }},
url = {https://tinyurl.com/LlamaIndex-Embedding},
note = {Accessed: 2024-08-07}
}

@www{LlamaIndex_Cosine_Similarity,
author = {LlamaIndex},
title = {{LlamaIndex Cosine Similarity }},
url = {https://tinyurl.com/LlamaIndex-Cosine-Similarity},
note = {Accessed: 2025-09-09}
}

@www{LlamaIndex_Vector_Store,
author = {LlamaIndex},
title = {{LlamaIndex Vector Store }},
url = {https://tinyurl.com/LlamaIndex-Vector-Store},
note = {Accessed: 2025-08-08}
}

@www{Elastic_Search_Relevance_Engine,
author = {Elastic Stack},
title = {{Elastic Search Relevance Engine}},
url = {https://www.elastic.co/elasticsearch/elasticsearch-relevance-engine},
note = {Accessed: 2025-08-08}
}

@www{Haystack,
author = {Haystack},
title = {{Haystack for Buidling RAG Piplelines}},
url = {https://haystack.deepset.ai/overview/intro},
note = {Accessed: 2025-08-08}
}

@www{META_FAISS,
author = {META FAISS},
title = {{META FAISS Vector Database}},
url = {https://faiss.ai/index.html},
note = {Accessed: 2025-08-08}
}

@www{Pinecone_Vector_Database,
author = {Pinecone},
title = {{Pinecone Vector Database}},
url = {https://www.pinecone.io/},
note = {Accessed: 2025-08-08}
}

@www{Chroma_Vector_Database,
author = {Chroma},
title = {{Chroma Vector Database}},
url = {https://docs.trychroma.com/},
note = {Accessed: 2025-08-08}
}

@www{OpenZeppelin_SmartContract,
author = {OpenZeppelin},
title = {{OpenZeppelin Smart Contract}},
url = {https://github.com/OpenZeppelin/openzeppelin-contracts},
note = {Accessed: 2025-08-09}
}

@www{Consensys_SmartContract,
author = {Consensys},
title = {{Consensys Smart Contract}},
url = {https://github.com/Consensys},
note = {Accessed: 2025-08-09}
}

@www{Zuora_Revenue_Sharing,
author = {Zuora},
title = {{Zuora Revenue Sharing}},
url = {https://www.zuora.com/},
note = {Accessed: 2025-08-09}
}

@www{Chargebee_Revenue_Sharing,
author = {Chargebee},
title = {{Chargebee Revenue Sharing}},
url = {https://www.chargebee.com/},
note = {Accessed: 2025-08-09}
}

@inproceedings{barnett2024seven,
  title={Seven failure points when engineering a retrieval augmented generation system},
  author={Barnett, Scott and Kurniawan, Stefanus and Thudumu, Srikanth and Brannelly, Zach and Abdelrazek, Mohamed},
  booktitle={Proceedings of the IEEE/ACM 3rd International Conference on AI Engineering-Software Engineering for AI},
  pages={194--199},
  year={2024}
}

@book{christian2016algorithms,
  title={Algorithms to live by: The computer science of human decisions},
  author={Christian, Brian and Griffiths, Tom},
  year={2016},
  publisher={Macmillan}
}

@article{chowdhery2023palm,
  title={Palm: Scaling language modeling with pathways},
  author={Chowdhery, Aakanksha and Narang, Sharan and Devlin, Jacob and Bosma, Maarten and Mishra, Gaurav and Roberts, Adam and Barham, Paul and Chung, Hyung Won and Sutton, Charles and Gehrmann, Sebastian and others},
  journal={Journal of Machine Learning Research},
  volume={24},
  number={240},
  pages={1--113},
  year={2023}
}

@inproceedings{xuretrieval,
  title={Retrieval meets Long Context Large Language Models},
  author={Xu, Peng and Ping, Wei and Wu, Xianchao and McAfee, Lawrence and Zhu, Chen and Liu, Zihan and Subramanian, Sandeep and Bakhturina, Evelina and Shoeybi, Mohammad and Catanzaro, Bryan},
  booktitle={The Twelfth International Conference on Learning Representations}
}

@article{roberts2020much,
  title={How much knowledge can you pack into the parameters of a language model?},
  author={Roberts, Adam and Raffel, Colin and Shazeer, Noam},
  journal={arXiv preprint arXiv:2002.08910},
  year={2020}
}

@article{ji2023survey,
  title={Survey of hallucination in natural language generation},
  author={Ji, Ziwei and Lee, Nayeon and Frieske, Rita and Yu, Tiezheng and Su, Dan and Xu, Yan and Ishii, Etsuko and Bang, Ye Jin and Madotto, Andrea and Fung, Pascale},
  journal={ACM Computing Surveys},
  volume={55},
  number={12},
  pages={1--38},
  year={2023},
  publisher={ACM New York, NY}
}

@article{menick2022teaching,
  title={Teaching language models to support answers with verified quotes},
  author={Menick, Jacob and Trebacz, Maja and Mikulik, Vladimir and Aslanides, John and Song, Francis and Chadwick, Martin and Glaese, Mia and Young, Susannah and Campbell-Gillingham, Lucy and Irving, Geoffrey and others},
  journal={arXiv preprint arXiv:2203.11147},
  year={2022}
}

@article{huang2023survey,
  title={A survey on hallucination in large language models: Principles, taxonomy, challenges, and open questions},
  author={Huang, Lei and Yu, Weijiang and Ma, Weitao and Zhong, Weihong and Feng, Zhangyin and Wang, Haotian and Chen, Qianglong and Peng, Weihua and Feng, Xiaocheng and Qin, Bing and others},
  journal={arXiv preprint arXiv:2311.05232},
  year={2023}
}

@inproceedings{zhang2024how,
title={How Language Model Hallucinations Can Snowball},
author={Muru Zhang and Ofir Press and William Merrill and Alisa Liu and Noah A. Smith},
booktitle={Forty-first International Conference on Machine Learning},
year={2024},
url={https://openreview.net/forum?id=FPlaQyAGHu}
}

@inproceedings{kuhn2023semantic,
title={Semantic Uncertainty: Linguistic Invariances for Uncertainty Estimation in Natural Language Generation},
author={Lorenz Kuhn and Yarin Gal and Sebastian Farquhar},
booktitle={The Eleventh International Conference on Learning Representations },
year={2023},
url={https://openreview.net/forum?id=VD-AYtP0dve}
}

@article{farquhar2024detecting,
  title={Detecting hallucinations in large language models using semantic entropy},
  author={Farquhar, Sebastian and Kossen, Jannik and Kuhn, Lorenz and Gal, Yarin},
  journal={Nature},
  volume={630},
  number={8017},
  pages={625--630},
  year={2024},
  publisher={Nature Publishing Group UK London}
}

@article{hron2024training,
  title={Training Language Models on the Knowledge Graph: Insights on Hallucinations and Their Detectability},
  author={Hron, Jiri and Culp, Laura and Elsayed, Gamaleldin and Liu, Rosanne and Adlam, Ben and Bileschi, Maxwell and Bohnet, Bernd and Co-Reyes, JD and Fiedel, Noah and Freeman, C Daniel and others},
  journal={arXiv preprint arXiv:2408.07852},
  year={2024}
}

@article{yang2024harnessing,
  title={Harnessing the power of llms in practice: A survey on chatgpt and beyond},
  author={Yang, Jingfeng and Jin, Hongye and Tang, Ruixiang and Han, Xiaotian and Feng, Qizhang and Jiang, Haoming and Zhong, Shaochen and Yin, Bing and Hu, Xia},
  journal={ACM Transactions on Knowledge Discovery from Data},
  volume={18},
  number={6},
  pages={1--32},
  year={2024},
  publisher={ACM New York, NY}
}

@www{vectara_price_plan,
author = {Vectara},
title = {{Vectara Price Plan}},
url = {https://vectara.com/pricing/},
note = {Accessed: 2024-09-01}
}

@www{nuclia_price_plan,
author = {Nuclia},
title = {{Nuclia Price Plan}},
url = {https://nuclia.com/pricing/},
note = {Accessed: 2024-09-01}
}

@www{llamaIndex_price_plan,
author = {Cloud LlamaIndex},
title = {{Cloud LlamaIndex Price Plan}},
url = {https://docs.cloud.llamaindex.ai/llamaparse/usage_data},
note = {Accessed: 2024-09-01}
}

@inproceedings{tang2025mba,
  title={MBA-RAG: a Bandit Approach for Adaptive Retrieval-Augmented Generation through Question Complexity},
  author={Tang, Xiaqiang and Gao, Qiang and Li, Jian and Du, Nan and Li, Qi and Xie, Sihong},
  booktitle={Proceedings of the 31st International Conference on Computational Linguistics},
  pages={3248--3254},
  year={2025}
}

@online{AWS2025,
  author = {Amazon},
  title = {AWS Pricing},
  url = {https://aws.amazon.com/pricing/},
  urldate = {2025-09-04},
  year = {2025}
}

@article{lin2025scorerag,
  title={ScoreRAG: A Retrieval-Augmented Generation Framework with Consistency-Relevance Scoring and Structured Summarization for News Generation},
  author={Lin, Pei-Yun and Tsai, Yen-lung},
  journal={arXiv preprint arXiv:2506.03704},
  year={2025}
}

@article{wang2024rear,
  title={Rear: A relevance-aware retrieval-augmented framework for open-domain question answering},
  author={Wang, Yuhao and Ren, Ruiyang and Li, Junyi and Zhao, Wayne Xin and Liu, Jing and Wen, Ji-Rong},
  journal={arXiv preprint arXiv:2402.17497},
  year={2024}
}

@www{LlamaIndex2025_Chunksize_Evaluation,
  author = {{LlamaIndex}},
  title = {{LlamaIndex Chunk Size Evaluation}},
  url = {https://www.llamaindex.ai/blog/evaluating-the-ideal-chunk-size-for-a-rag-system-using-llamaindex-6207e5d3fec5},
  urldate = {2025-06-09},
  year = {2025}
}

@article{belgacem2022dynamic,
  title={Dynamic resource allocation in cloud computing: analysis and taxonomies},
  author={Belgacem, Ali},
  journal={Computing},
  volume={104},
  number={3},
  pages={681--710},
  year={2022},
  publisher={Springer}
}

@article{ghasemi2018cost,
  title={A cost-aware mechanism for optimized resource provisioning in cloud computing},
  author={Ghasemi, Safiye and Meybodi, Mohammad Reza and Fooladi, Mehdi Dehghan Takht and Rahmani, Amir Masoud},
  journal={Cluster Computing},
  volume={21},
  number={2},
  pages={1381--1394},
  year={2018},
  publisher={Springer}
}

@article{kumar2020autonomic,
  title={Autonomic cloud resource provisioning and scheduling using meta-heuristic algorithm},
  author={Kumar, Mohit and Sharma, Subhash Chander and Goel, Shalini and Mishra, Sambit Kumar and Husain, Akhtar},
  journal={Neural Computing and Applications},
  volume={32},
  number={24},
  pages={18285--18303},
  year={2020},
  publisher={Springer}
}

@inproceedings{younis2024comprehensive,
  title={A comprehensive analysis of cloud service models: IaaS, PaaS, and SaaS in the context of emerging technologies and trend},
  author={Younis, Rehmana and Iqbal, Mansoor and Munir, Khalid and Javed, Muhammad Aaqib and Haris, Muhammad and Alahmari, Saad},
  booktitle={2024 international conference on electrical, communication and computer engineering (ICECCE)},
  pages={1--6},
  year={2024},
  organization={IEEE}
}

@www{LlamaIndex2025_similarity_metrics,
  author = {LlamaIndex},
  title = {LlamaIndex Similarity Metrics},
  url = {https://tinyurl.com/m4p9ucus},
  urldate = {2025-17-09},
  year = {2025}
}

@www{cosine_similarity_effectiveness,
  author = {ScienceDirect},
  title = {LlamaIndex Similarity Metrics},
  url = {https://www.sciencedirect.com/topics/computer-science/cosine-similarity},
  urldate = {2025-17-09},
  year = {2025}
}

@inproceedings{sarthi2024raptor,
  title={Raptor: Recursive abstractive processing for tree-organized retrieval},
  author={Sarthi, Parth and Abdullah, Salman and Tuli, Aditi and Khanna, Shubh and Goldie, Anna and Manning, Christopher D},
  booktitle={The Twelfth International Conference on Learning Representations},
  year={2024}
}

@inproceedings{li2024retrieval,
  title={Retrieval Augmented Generation or Long-Context LLMs? A Comprehensive Study and Hybrid Approach},
  author={Li, Zhuowan and Li, Cheng and Zhang, Mingyang and Mei, Qiaozhu and Bendersky, Michael},
  booktitle={Proceedings of the 2024 Conference on Empirical Methods in Natural Language Processing: Industry Track},
  pages={881--893},
  year={2024}
}

\begin{IEEEbiography}[{\includegraphics[width=1in,height=1.25in,clip,keepaspectratio]{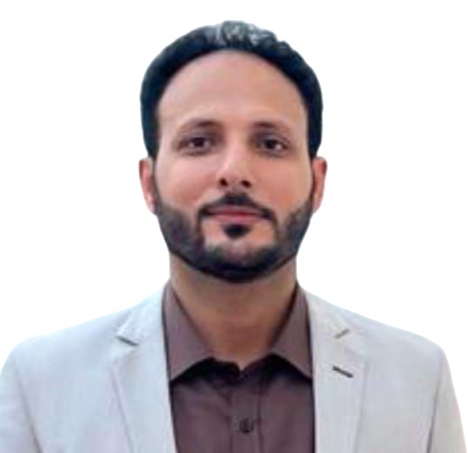}}]{Shawqi Al-Maliki}

is currently a postdoctoral researcher at Hamad Bin Khalifa University (HBKU), Doha, Qatar. He received his M.Sc. degree from King Fahd University of Petroleum and Minerals (KFUPM) in Dhahran, Saudi Arabia, in 2018, and his Ph.D. degree in computer science from HBKU in 2023. 
His research interests span adversarial robustness, LLMs security, privacy, reliability, and human-LLM alignment. Prior to his academic career, Dr. Al-Maliki held positions as a Full-Stack Web Developer \& Project Manager at Alyaum Media House (2009-2019) and a computer instructor at New Horizon Institute (2006-2008), both in Saudi Arabia.
\end{IEEEbiography}

\begin{IEEEbiography}[{\includegraphics[width=1in,height=1.25in,clip,keepaspectratio]{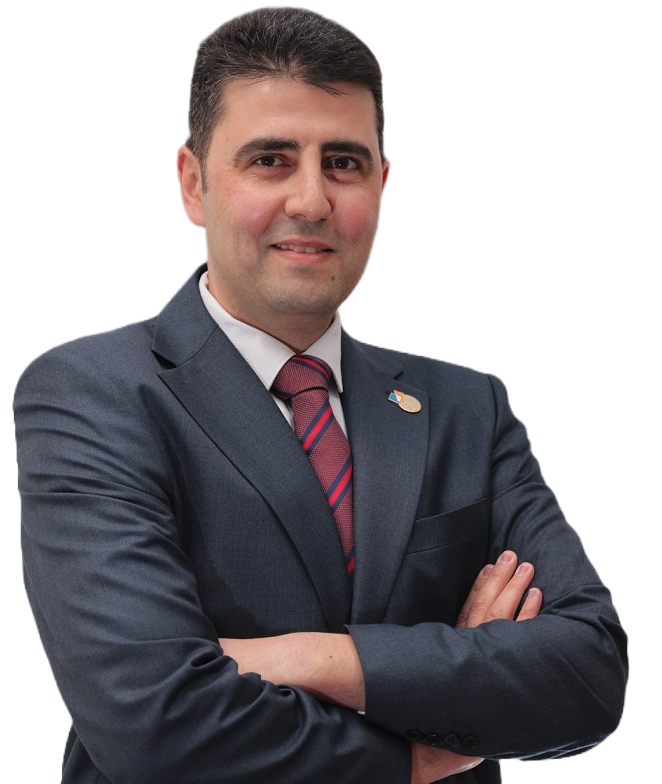}}]{Dr. Ammar Gharaibeh}
 is an Associate Professor with the Computer Engineering department of German Jordanian University, and currently serving as the Acting Dean of the School of Computing at GJU. He received his Ph.D. degree in Computer Engineering from New Jersey Institute of Technology, his M.S. degrees in Computer Engineering from Texas A\&M University in 2009, and his B.S. degree with honors in Electrical Engineering from Jordan University of Science \& Technology in 2006. His research interests span the areas of Online Algorithms, Wireless Networks, and AI applications in Wireless Sensor Networks. 
\end{IEEEbiography}

\begin{IEEEbiography}[{\includegraphics[width=1in,height=1.25in,clip,keepaspectratio]{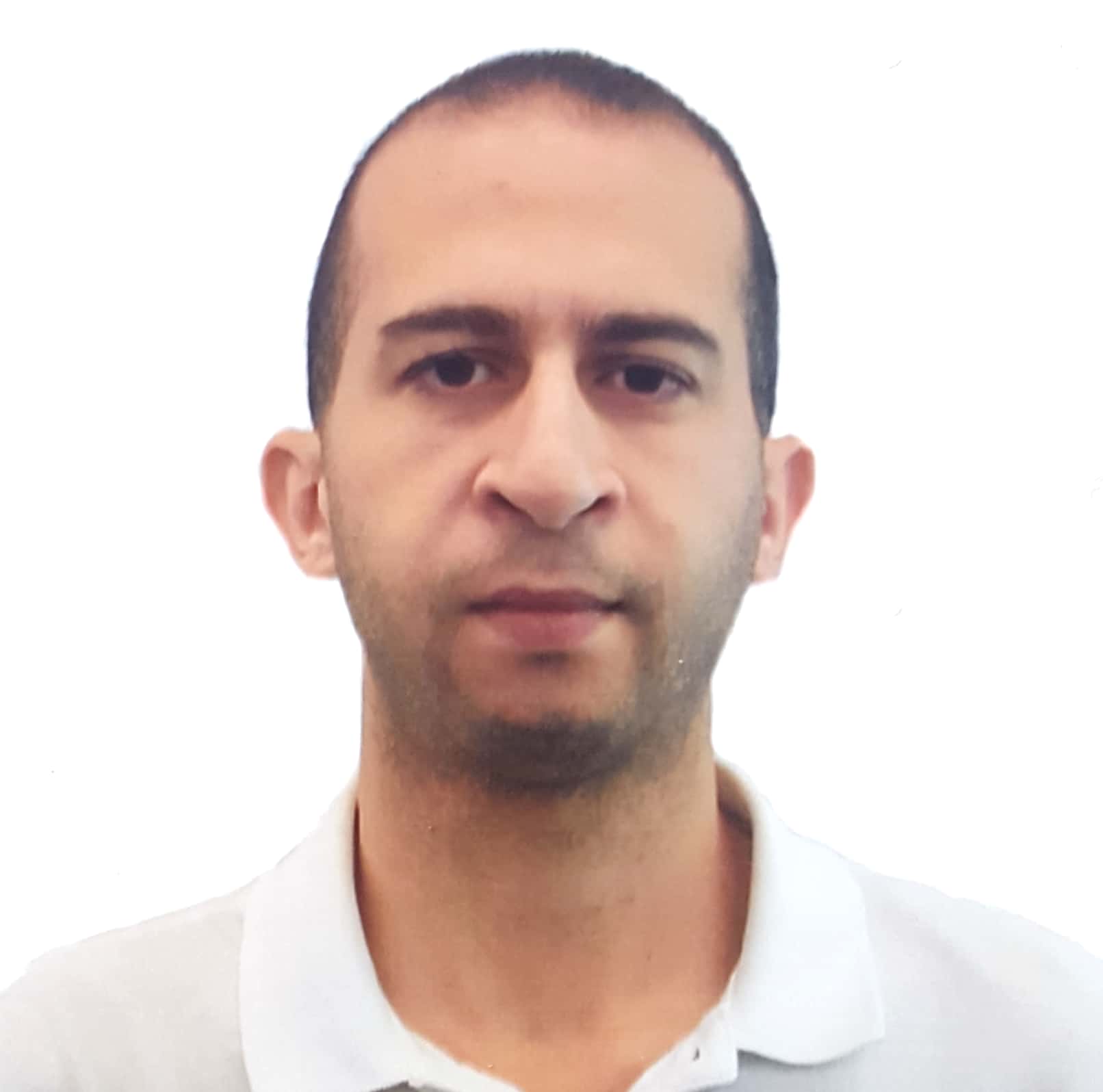}}]{Mohamed Rahouti }
received his M.S. degree in Statistics and Ph.D. degree in Electrical Engineering from the University of South Florida, Tampa, FL, USA, in 2016 and 2020, respectively. He is currently an Assistant Professor with the Department of Computer and Information Science, Fordham University, New York, NY, USA. His research was supported by the National Science Foundation (NSF), the Florida Center for Cybersecurity, and the Qatar Research Development and Innovation Council. His current research focuses on computer networking and security, blockchain technology, and artificial intelligence/machine learning.
\end{IEEEbiography}

\begin{IEEEbiography}[{\includegraphics[width=1in,height=1.25in,clip,keepaspectratio]{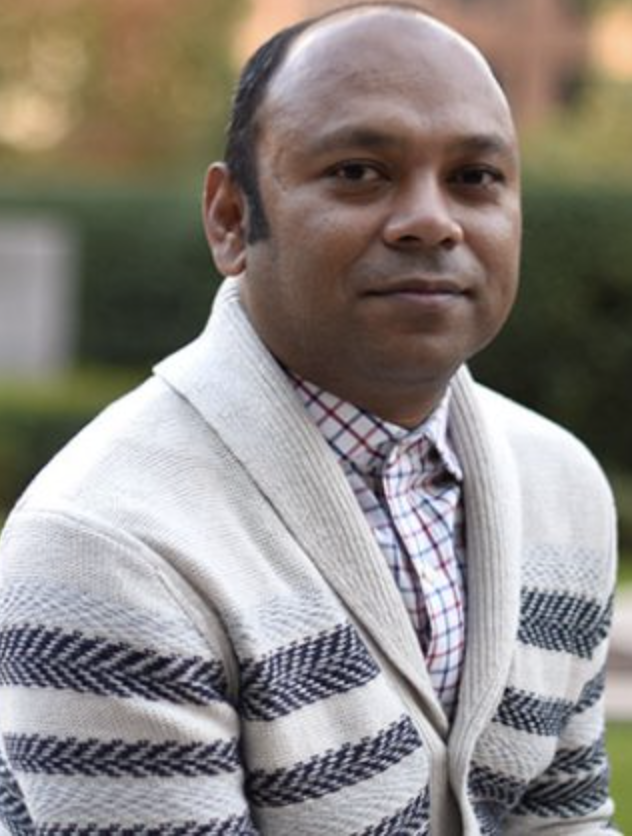}}]{Ruhul Amin }
is an Assistant Professor in the Department of Computer and Information Science at Fordham University. He earned his PhD in Computer Science from Stony Brook University, where he was advised by the distinguished Professor Steven Skiena.
Prior to his current role, Ruhul served as a Graduate Research Assistant at the Institute of Advanced Computational Science (IACS), where he led multiple collaborative research initiatives in Bioinformatics and Social Science. He also brings industry experience from his time with the Amazon AWS AI Algorithms group, where he developed deep learning-based algorithms for anomaly detection.
\end{IEEEbiography}

\begin{IEEEbiography}[{\includegraphics[width=1in,height=1.25in,clip,keepaspectratio]{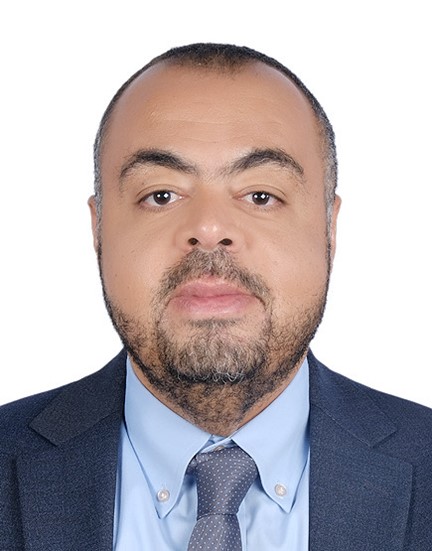}}]{Dr. Mohamed Abdallah}
is currently a Professor and Associate Dean of Undergraduate Studies and Quality Assurance at the College of Science and Engineering, Hamad Bin Khalifa University (HBKU). He received his M.Sc. and Ph.D. degrees from the University of Maryland at College Park, USA in 2001 and 2006, respectively.
His research interests lie in AI for communications, with a focus on 6G wireless networks, wireless security, electric vehicles, and smart grids. He has authored more than 220 journal and conference papers, contributed to four book chapters, and co-invented four patents.
Dr. Abdallah has received several prestigious awards, including the Research Fellow Excellence Award at Texas A\&M University at Qatar in 2016, and Best Paper Awards at multiple IEEE conferences, such as IEEE BlackSeaCom 2019 and the IEEE First Workshop on Smart Grid and Renewable Energy in 2015. Additionally, he was a recipient of the Nortel Networks Industrial Fellowship for five consecutive years (1999–2003).

\end{IEEEbiography}

\begin{IEEEbiography}[{\includegraphics[width=1in,height=1.25in,clip,keepaspectratio]{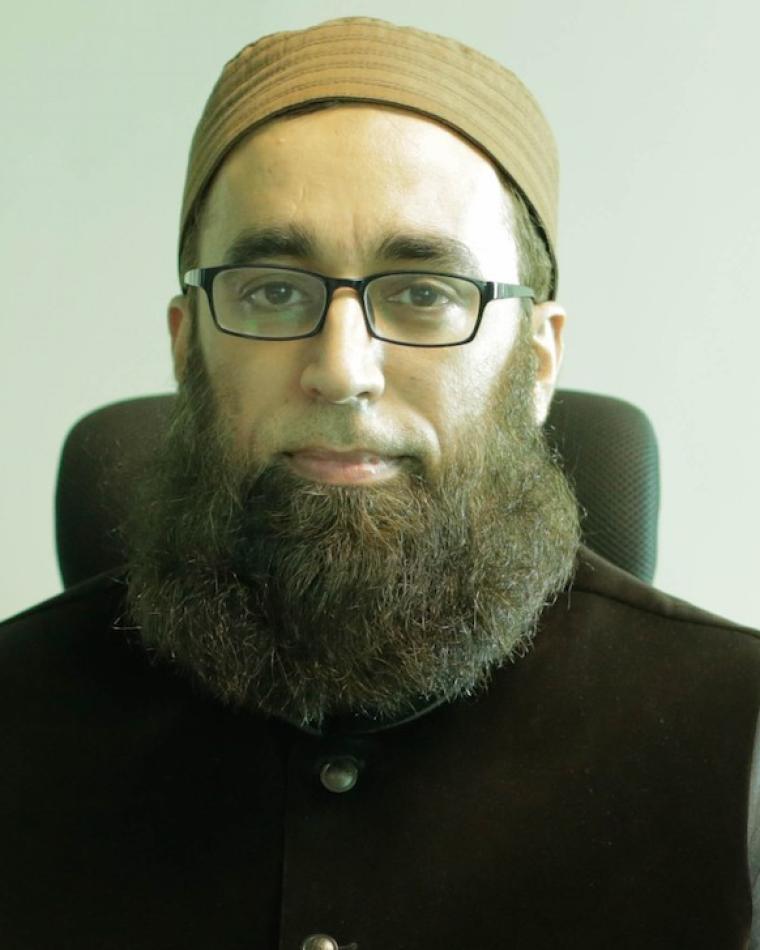}}]{Junaid Qadir }
 is a Professor of Computer Engineering at Qatar University in Doha, Qatar. His core research interests lie in human-centered artificial intelligence, AI ethics, and the application of AI in education and healthcare. His research is widely published, with more than 250 peer-reviewed articles, including in leading journals such as IEEE Transactions on Pattern Analysis and Machine Intelligence, IEEE Transactions on Artificial Intelligence, and the Journal of Artificial Intelligence Research. He has won a range of awards, including the IEEE Engineering in Medicine and Biology Prize Paper Award (2023), the Emerald Literati Outstanding Paper Award (2023), and the Qatar University College of Engineering Dean’s Awards for Excellence in Research (2024) and Service (2025). 
 He has secured competitive research funding from Facebook Research, the Qatar National Research Fund, and the Higher Education Commission of Pakistan. He is an ACM Distinguished Speaker (2020–2026), an IEEE Distinguished Lecturer (2025–2026) of the IEEE Education Society, a Senior Member of IEEE since 2014, and a Senior Member of ACM since 2020.
\end{IEEEbiography}

\begin{IEEEbiography}[{\includegraphics[width=1in,height=1.25in,clip,keepaspectratio]{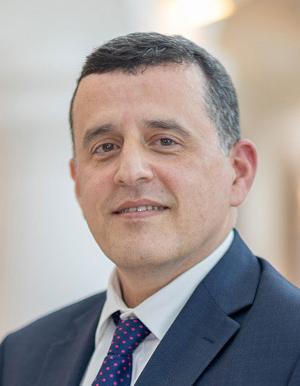}}]{Prof. Ala Al-Fuqaha}
is the Associate Provost for Teaching and Learning at Hamad Bin Khalifa University (HBKU). He is a professor at the College of Science and Engineering, HBKU. Before HBKU, he was Professor and director of the NEST research lab at the Computer Science Department of Western Michigan University (WMU). His research interests include the use of machine learning in general and deep learning in particular in support of the data-driven and self-driven management of large-scale deployments of IoT and smart city infrastructure and services, and management and planning of software-defined networks (SDN). He is a senior member of the IEEE, a senior member of the ACM, and an ABET Program Evaluator (PEV) and commissioner. He served on editorial boards of multiple journals, including IEEE Communications Letters, IEEE Network Magazine, and Springer AJSE. He also served as chair, co-chair, and technical program committee member of multiple international conferences, including IEEE VTC, IEEE Globecom, IEEE ICC, and IWCMC. 
\end{IEEEbiography}

\end{document}